\documentclass[10pt,righttag]{siamltex}
\usepackage{amsmath,amssymb,fontenc}
\usepackage{a4wide}
\usepackage[dvips]{graphicx}
\usepackage{psfrag}
\usepackage{epsfig,subfigure}
\usepackage{amscd}
\usepackage{dsfont}
\usepackage{amscd}
\usepackage{algorithm,algorithmic}
\usepackage{setspace}

\usepackage[latin1]{inputenc}

\usepackage{color}

\DeclareMathOperator*{\Argmax}{argmax}

\title{Stochastic Algorithm
      For Bayesian Mixture Effect Template  Estimation }

\author{
S. Allassonni�re$^1$, E. Kuhn$^2$
}
\begin{document}
\maketitle

 \begin{center}
    CMAP - Ecole Polytechnique$^1$  \hspace{1cm} LAGA - Universit�
    Paris 13$^2$\\
    Route de Saclay \hspace{3cm} 99, Av. J.-B. Cl�ment\\
91128 Palaiseau, France  \hspace{2cm} 93430 Villetaneuse, France
      \end{center}


\newcommand{\nin}{\in\!\!\!\!\!/ \: \,}
\newcommand{\C}{\mathbb{C}}
\newcommand{\V}{\mathbb{V}}
\newcommand{\R}{\mathbb{R}}
\newcommand{\Z}{\mathbb{Z}}
\newcommand{\N}{\mathbb{N}}
\newcommand{\Q}{\mathbb{Q}}
\newcommand{\M}{\mathcal{M}}
\newcommand{\T}{\mathcal{T}}
\newcommand{\s}{\mathcal{S}}
\newcommand{\G}{\Gamma}
\newcommand{\e}{\eta}
\newcommand{\m}{\ell}
\newcommand{\la}{\boldsymbol{\tau}}
\newcommand{\te}{\theta}
\newcommand{\si}{\sigma}
\newcommand{\Te}{\Theta}
\newcommand{\Sa}{\mathcal{S}_a}
\newcommand{\x}{v}
\newcommand{\ot}{\tau}
\newcommand{\tS}{S}
\newcommand{\tntrans}{\tilde{\Pi}}
\newtheorem{notation}{Notation}
\newtheorem{Def}{Definition}
\newtheorem{prop}{Proposition}
\newtheorem{rem}{Remark}
\newtheorem{Cor}{Corollary}

\newcommand{\Kp}{\mathbf{K_p}}
\newcommand{\Kpb}{K_{p}^\beta}
\newcommand{\Kpbi}{K_{p}^{\beta_i}}
\newcommand{\Kg}{\mathbf{K_g}}
\newcommand{\fm}{\frak{m}}
\newcommand{\gm}{\gamma_{\fm}}
\newcommand{\Syme}{\text{Sym}_{2k_g,*}^+(\mathbb{R})}
\newcommand{\Sym}{\Sigma_{2k_g}^+(\mathbb{R})}
\newcommand{\atan}{\text{atan}}
\newcommand{\tT}{\tilde{\Theta}}
\newcommand{\otT}{\overline{\tT}}
\newcommand{\htR}{\hat{\theta}^R}
\newcommand{\htRn}{\hat{\theta}^R_n}
\newcommand{\dimb}{\text{dim}_\beta}
\newcommand{\dima}{\text{dim}_\alpha}
\newcommand{\dimy}{\text{dim}_y}
\newcommand{\Ia}{\mathcal{I}_R}
\newcommand{\Ial}{\mathcal{I}_R^l}

\newcommand{\lstat}{\textbf{$\pi$}}
\newcommand{\ntrans}{\textbf{$\Pi$}}
\newcommand{\cand}{p}
\newcommand{\compdens}{f}
\newcommand{\densMC}{f}
\newcommand{\smallset}{\texttt{C}}

\newcommand{\nmck}{J_k}
\newcommand{\poids}{p_{\nmc,\e}(t|\xi_{i},y_i)}

\newcommand{\bfe}{\boldsymbol{\varepsilon}}
\newcommand{\bfr}{\boldsymbol{\rho}}
\newcommand{\bfo}{\boldsymbol{1}}
\newcommand{\bdbeta}{\boldsymbol\beta}
\newcommand{\bdtau}{\boldsymbol\tau}
\newcommand{\dy}{y}
\newcommand{\bdy}{\boldsymbol{\dy}}
\newcommand{\z}{z}
\newcommand{\bdz}{\boldsymbol{\z}}
\def\pixel{u}
\def\locpixel{v}
\newcommand{\nmc}{J}

\def\Gammap{\Sigma_p}
\def\Gammag{\Sigma_g}
\def\I0{I_\alpha}
\def\Igi{I_{\alpha_i}}
\def\pk{\x_{p,k}}
\def\pk{\locpixel_{p,j}}
\def\pkp{\locpixel_{p,j'}}
\def\gk{\locpixel_{g,j}}
\def\gkp{\locpixel_{g,j'}}
\def\K{{K}}
\def\mm{{\eta}}
\def\tt{\tau_m}
\def�{\mathbb{P}}
\def\T{\mathcal{T}}
\def\M{\mathcal{H}}
\def\E{\mathcal{E}}
\def\F{\mathcal{F}}
\def\B{\mathcal{B}}
\def\Kapa{\mathcal{K}}
\def\kapa{\mathds{k}}

\newcommand{\qobs}{q_{obs}}
\newcommand{\qmiss}{q_{m}}
\newcommand{\qcond}{q_c}
\newcommand{\qcomp}{q}
\newcommand{\qbayes}{q_B}
\newcommand{\qpara}{q_{para}}
\newcommand{\qj}{q_{j}}
\newcommand{\qpost}{q}
\newcommand{\betamj}{\beta^{-j}} 
\newcommand{\bdbetamj}{\boldsymbol{\beta}^{-j}}
\newcommand{\betajb}{\boldsymbol{\beta}_{b\to j}} 

\begin{abstract}
The estimation of  probabilistic deformable template models in
computer vision or of probabilistic atlases in Computational Anatomy
are core issues in both fields.
A first coherent statistical framework where the geometrical variability is
modelled as a hidden
random variable has been
 given by Allassonniere, Amit and Trouve in \cite{AAT}. They introduce
 a Bayesian approach and
 mixture of them to estimate deformable template models.
A consistent stochastic algorithm has been introduced in
\cite{aktdefmod} to face the problem encountered in \cite{AAT} for the
convergence of the estimation algorithm for the one component model in
the presence of noise.
We propose here to go on in this direction of using some ``SAEM-like''
algorithm to approximate the MAP estimator in the general Bayesian setting of
mixture of deformable template models.
We also prove the convergence of our algorithm toward a critical
point of the penalised likelihood of the observations  and
illustrate this with handwritten digit images and medical images.
\end{abstract}

\textbf{keywords}
 stochastic approximations, non rigid-deformable templates,
 shapes statistics, MAP estimation, Bayesian method, mixture models.\\

 \textbf{AMS}
 60J22, 62F10, 62F15, 62M40.


\section{Introduction}
The issue of 
 representing and analysing some geometrical structures
upon which some deformations can act is a challenging question in
applied mathematics
as well as 
 in Computational Anatomy.
One central point is the modelisation of varying objects, and
the quantification of this variability with respect to one or several
reference models which will be called templates. This is known as
``Deformable Templates'' \cite{gre93}.
 To our best knowledge, the problem of constructing probabilistic
models of variable shapes in order to statistically quantify
this variability has not been  successfully addressed
 yet in spite of  its importance. For example, modelling the
 anatomical variability of organs around an ideal shape is of a crucial
 interest in the medical domain in order to find some characteristic
 differences between populations (pathological and control), or to exhibit
 some pathological kind of deformations or shapes of an organ.

Many solutions have been proposed to face the problem of the template
definition. They go from some generalised Procruste's means with a
variational \cite{Glaunes:MFCA:06} or statistical \cite{glasbey01}
point of view to some statistical models like Active Appearance Model
\cite{cootes98:_activ_appear} or Minimum Description Length methods
\cite{marsland07}.
Unfortunately, all these methods only focus on the template whereas
the geometrical variability is computed afterwards (using
PCA). This contradicts with the fact that a metric is required
to compute the template through the
computation of deformations.
Moreover, they do not really differ from the variational point
of view since they consider the deformations as some nuisance
parameters which have to be estimated  and not as some
unobserved random variables.

The main goal of this paper is to propose a coherent estimation of both
photometric model and
geometrical distribution in a given population.
Another issue addressed here is the clustering problem.
Given a set
of images, the statistical estimation of the component weights and of the
image labels is usually supervised, at least the number of components
is fixed. The templates of each component and the label are estimated
iteratively (for example in methods like  K-means) but the
geometry, and related to this the metric used to compute the distances
between elements, is still fixed. Moreover, the label,
which is not observed, is, as the deformations, considered as a
parameter and not as a hidden random variable.
These methods do not lead to a statistical coherent framework for the
understanding of deformable template estimation and all these
iterative algorithms derived from those approaches do
not have a statistical interpretation as the parameter optimisation
 of a generative model describing the data.

In this paper we consider the statistical framework for dense
deformable templates developed by Allassonniere, Amit and Trouve in
\cite{AAT} in the generalised case of mixture model for multicomponent
estimation.
 Each image taken from a database is supposed to be generated from a
 noisy random deformation of a 
  template image picked randomly among
 a given set of possible templates. All the templates are assumed to
 be drawn from a
 common prior distribution on the template image space. To propose a
 generative model, each deformation and each image label
  have to be
 considered as \textit{hidden} variables. The templates, the parameters of
 the deformation laws and the components weights are the parameters of
 interest.
 This generative model allows to automatically decompose the
 database into components and, at the same time, estimates the parameters
 corresponding to each component while increasing the likelihood of the
 observations.

 Given this parametric statistical Bayesian model,
 the parameter estimation is performed in \cite{AAT} by a  Maximum
A Posteriori (MAP). 
The authors  carry  out this estimation problem using a
deterministic and
iterative scheme based on the EM (Expectation Maximisation) algorithm
where the posterior distribution is approximated by a Dirac measure on
its mode.
Unfortunately, this gives an algorithm whose convergence toward
 the MAP estimator cannot be proved.
Moreover, as shown by the experiments in that paper, the convergence
is lost within a noisy setting.

 Our goal in this paper is to propose some stochastic iterative method
 to reach the MAP estimator for which we will be able to get a
 convergence result as already done for the one component case in
 \cite{aktdefmod}.
We propose  to use a stochastic version of
the EM algorithm to reach the maximum of the posterior distribution.
We use the Stochastic Approximation EM (SAEM) algorithm introduced by
Delyon et al in
\cite{DLM} coupled with a Monte Carlo Markov Chain (MCMC)
method. This coupling algorithm has been introduced by Kuhn  and
Lavielle in \cite{kuhnlavielle} in the case where the missing
variables had a compact support. Contrary to the one component model
where we can couple the iteration of the SAEM algorithm with the
Markov chain evolution (cf. \cite{aktdefmod}), we show
here that it cannot be driven numerically. We need to consider
an alternative method. We propose to simulate  the
hidden variables using some auxiliary Markov chains,
one per component, to approach the posterior distribution.
We prove the convergence of our algorithm
for a non compact setting by adapting
Delyon's theorem about general stochastic approximations and
introducing truncation on random boundaries as 
in
\cite{andrieumoulinespriouret}.

The paper is organised as follows: in Section \ref{obs} we first
recall the observation mixture model proposed by Allassonniere, Amit and
Trouve in \cite{AAT}. In Section \ref{SAEM},  we describe  the
stochastic algorithm used in our
particular setting. Section \ref{convergence} is devoted to the
convergence theorem. Experiments on 2D  real data sets are presented
in Section \ref{experiments}. The proofs of the convergence of the
algorithm are postponed in  Section \ref{proof} whereas Conclusion
and Discussion are given in Section \ref{conclusion}.

\section{The observation model}\label{obs}
We consider the framework of multicomponent model introduced in \cite{AAT}.
Given a sample of gray level images $(y_i)_{1\leq i\leq n}$ observed on
a grid of pixels $\{\x_u \in D \subset \R^2, u \in \Lambda\}$ where $D$
is a continuous domain and $ \Lambda$ the pixel network, we are
looking for some
 template images which
explain the population. Each of these images is a real function $I_0
: \R^2 \to
 \R$  defined on the whole plane. An
observation $y$ is supposed to be a discretisation on $\Lambda $ of a
deformation of one of the templates plus an independent
additive noise. This leads to assume the existence of an unobserved
deformation field $z:\R^2 \to
\R^2$ such that for $u \in \Lambda$ :
\begin{eqnarray*}
y(u) = I_0(\x_u - z(\x_u)) + \epsilon (u)\,,
\end{eqnarray*}
where $ \epsilon$ denotes an additive noise.

\subsection{Models for templates and deformations}
We use the same framework as chosen in \cite{AAT} to describe both the
templates  $I_0$ and
the deformation fields $z$.
Our model takes into account two complementary sides: photometric
-indexed by $p$- corresponding to the templates and the noise
variances, and geometric -indexed by $g$- corresponding to the
deformations.
The templates $I_0$ and the deformations $z$ are assumed to belong to
 some finite dimensional subspaces of two reproducing kernels
Hilbert spaces $V_p$ and $V_g$ (determined by their respective kernels
$K_p$  and $K_g$). We choose a representation of both of them by finite
linear combinations of the
kernels centred
at some fixed landmark points in the domain $D$: $(\pk)_{1\leq j\leq
  k_p}$
respectively
$(\gk)_{1\leq j\leq k_g}$.
 They are therefore parametrised by the
coefficients
$\alpha \in \R^{k_p}$ and $\beta  \in
(\R^{k_g})^2 $ which yield: $\forall \x \in D $,
\begin{eqnarray*}\label{template}
\I0(\x) &\triangleq& (\Kp \alpha)(\x) \triangleq \sum\limits_{j=1}^{k_p}
K_p (\x,\pk) \alpha^j \,,\\
z_\beta(\x) &\triangleq& (\Kg \beta)(\x) \triangleq \sum\limits_{j=1}^{k_g} K_g (\x,\gk)
\beta^j.
\end{eqnarray*}

\subsection{Parametrical model} \label{model}

In this paper, we consider  a mixture of the deformable template
models  which
allows a fixed
number $\tau_m$ of components in each training set. This means that the
data will be separated in  $\tau_m$ (at most) different
components by the algorithm.

Therefore, for each observation $y_i$, we consider the pair
$(\beta_i, \tau_i)$ of unobserved variables which correspond
respectively to
the deformation field and to the label of image $i$.
We denote below by $\bdy^t\triangleq (y_1^t,\ldots,y_n^t)$, by
$\bdbeta^t\triangleq (\beta_1^t,\ldots , \beta_n^t )$ and by $\bdtau^t\triangleq  (\tau_1,\ldots,\tau_n ) $. The
generative model is:
\begin{equation} \label{eqmodel}
\left\{
\begin{array}{l}
\bdtau \sim \otimes _{i=1}^n \sum\limits_{t =1}^{\tt} \rho_{t}
\delta_{t} \ | \ (\rho_t )_{ 1\leq t \leq \tau_m}  \,,\\
\\
\displaystyle{\bdbeta \sim \otimes_{i=1}^n \mathcal{N}
(0,\Gamma_{g,\tau_i}) \ |  \   \ \bdtau, \ (\Gamma_{g,t})_{1\leq t \leq \tau_m  } } \,, \\
\\
\displaystyle{\bdy \sim  \otimes_{i=1}^n \mathcal{N} (z_{\beta_i}
  I_{\alpha_{\tau_i}},
\sigma_{\tau_i}^2 Id_{| \Lambda |} ) \ | \ \bdbeta, \  \bdtau}, \ (\alpha_t,\si^2_t)_{1\leq t
\leq \tau_m} \,,
\end{array} \right.
\end{equation}
where $z_\beta I_\alpha(u) = I_\alpha(\x_u - z_\beta (\x_u))$ is the action of the
deformation on the template $I_\alpha$, for $u$ in $\Lambda$ and $ \delta_t $
is the Dirac function on $t$.
The parameters of interest are the vectors $(\alpha_t)_{1\leq t\leq \tau_m}$ coding
the templates, the variances  $(\sigma_t^2)_{1\leq t\leq
  \tau_m}$
of the additive noises, the covariance matrices
$(\Gamma_{g,t} )_{1\leq t\leq \tau_m}$ of the deformation fields and the component weights $
(\rho_t)_{1\leq t\leq \tau_m}$.
We denote by $(\theta_t,\rho_t)_{1\leq t\leq \tau_m} $ the parameters so that
 $\theta_{t}$ corresponds to the parameters composed of the
photometric
part $(\alpha_t, \sigma^2_{t})$ and the geometric part $\Gamma_{g,t}$ for component $t$.
We assume that for all $1\leq t \leq \tau_m $, the parameter $\theta_t=(\alpha_t, \sigma_t^2,
\Gamma_{g,t})$ belongs
to the open space $\Theta$ defined as
$  \Theta =\{\ (\alpha,\sigma^2,\Gamma_g)\ |\
\alpha\in\mathbb{R}^{k_p}, \, |\alpha|< R, \ \sigma>0,\ \Gamma_g \in \Syme\ \} \,,
$
where $R$ is an arbitrary  positive constant and $\Syme$ is the set of
strictly positive symmetric
matrices. Concerning the weights $(\rho_t)_{1\leq t\leq \tau_m}$, we
assume that they belong to the set
$ \varrho=\left\{(\rho_t)_{1\leq t\leq \tau_m} \in ]0,1[^{\tau_m} \ | \
\sum\limits_{t=1}^{\tau_m} \rho_t=1\right\} \, .
$\\

\begin{rem}
This yields a generative model: given the parameters of the model, to
get a realisation of an image, we
first draw a label $\tau$ with respect to the probability law $
\sum\limits_{t=1}^{\tau_m}  \rho_{t}
\delta_{t}$. Then, we simulate a deformation field $\beta$ using the covariance
matrix corresponding to component $\tau$ according to $\mathcal{N}
(0,\Gamma_{g,\tau})$. We apply it to the template of the $\tau^{th}$
component. Last, we add an independent Gaussian noise of variance
$\si^2_\tau$.   \\
\end{rem}

We choose a normal distribution for the unobserved deformation
variable because of the
background we have in image analysis. Indeed,
the registration problem is an
issue that has been studied deeply for the past two decades. The goal
is, given two images, to find the best deformation that will match one
image close to the other. Such methods require to choose the kind of
deformations that are allowed (smooth, diffeomorphic, etc).
These conditions are equivalent, for
some of these methods, to choose a covariance matrix that enables to
define an inner product between two deformations coded by a vector $\beta$
(cf. \cite{mty02,agp91}). The
regularisation term of the matching energy in the
small deformation framework treated in this paper can be
written as~: $ \beta^t \Gamma^{-1}_g \beta$. This looks like the logarithm of the
density of a 
Gaussian distribution on $\beta$ with $0$ mean and a covariance matrix
$\Gamma_g$. The link between these two points of view has been
given in \cite{AAT}; the mode of the posterior distribution
equals the solution of a general matching problem.
This is why we therefore set on the deformation vector $\beta$ such a
distribution.
Moreover, many experiments have been run using a large variety of such
a matrix which
gives us now a good initial guess for our parameter. This leads us to
consider a Bayesian approach with  a weakly
informative prior.

\subsection{The Bayesian approach}\label{bayes}

The information given by the image analysis background is here introduced
mathematically in terms of prior laws on the parameters of  model
(\ref{eqmodel}).
As already mentioned in the previous paragraph, this background
knowledge
enables to determine a good initial guess for the laws and the values
of the hyper-parameters.
As well as for the covariance matrix $\Gamma_g$, the same arguments
are true for the noise variance $\si^2$. In the
registration viewpoint, this variance is the tradeoff between the
deformation cost and the data attachment term that compose the energy
to minimise. An empirical  good initial guess is therefore known
as well.

On another hand, the high dimensionality of the parameters can lead to
degenerated maximum likelihood
estimator when the training sample is small.
While introducing  prior distributions, the estimation with small samples is
still possible. The importance of these prior distributions in the
estimation problem
has been shown in \cite{AAT}.
The solution of the estimation equation can be
interpreted as barycenters between the hyper-parameters of the priors
and the empirical values.
This ensures easy computations and other theoretical properties as for example, the invertibility of
the covariance matrix
$\Gamma_g$.
The role of the other hyper-parameters are discussed in the
experiments.\\

We use a generative model which includes natural standard conjugate prior
distributions with
\textit{fixed} hyper-parameters. These distributions are
an  inverse-Wishart priors on
each $\Gamma_{g,t}$ and  $\si_t^2$ and  a normal prior on each $\alpha_t$, for
all $1\leq t \leq \tau_m$.
All priors are assumed independent.
 Then,
\begin{equation*}\label{priors} \left\{
\begin{array}{l}
\displaystyle{\nu_p(d\alpha, d\sigma^2) \varpropto
\exp \left(-\frac{1}{2} (\alpha-\mu_p)^t (\Gammap)^{-1} (\alpha - \mu_p) \right)
\left(
\exp\left(-\frac{\sigma_0^2}{2 \sigma^2}\right) \frac{1}{\sqrt{\sigma^2}}
\right)^{a_p}
  d\sigma^2
 d\alpha}, \ a_p \geq 3 \,,\\
\displaystyle{\nu_g(d\Gamma_g) \varpropto \left( \exp(-\langle
 \Gamma_g^{-1} ,
 \Gammag \rangle_F /2) \frac{1}{\sqrt{|\Gamma_g|}}  \right)^{a_g}
 d\Gamma_g},\ a_g\geq 4k_g+1\,,
\,
\end{array} \right.
\end{equation*}
where $\langle A,B
\rangle_F \triangleq tr(A^tB)$ is the scalar product of two matrices $A$ and $B$ and
$tr$ stands for the trace.

For the prior law $\nu_\rho$, we choose
the Dirichlet distribution, $\mathcal{D}(a_\rho) $, with density $$ \nu_\rho(\rho)
\propto \left( \prod\limits_{t=1}^{\tt}
\rho_t \right) ^{a_\rho}, \text{ with fixed parameter }
a_\rho\,.$$

\section{Parameter estimation using a stochastic
  version of the  EM   algorithm}\label{SAEM}

For the sake of simplicity, let us denote by $N \triangleq 2nk_g$ and by  $\T\triangleq
\{1,\dots,\tau_m\}^n$
so that the missing deformation variables take their
values in $\R^N$ and the missing labels in $\T$.
We also introduce the following notations:
$\mm =(\theta,\rho) \text{ with }
\theta=(\theta_{t} )_{1 \leq
t \leq \tau_m}
\text{ and }
\rho= \left(\rho_t \right)_{1 \leq t \leq \tau_m} \, .
$
\\

In our Bayesian framework, we
choose the MAP estimator  to estimate the parameters:

 \begin{eqnarray} \label{MAP}
 \tilde{\eta}_n = \Argmax_{\eta} \qbayes(\eta|\bdy)\,,
 \end{eqnarray}
 where $\qbayes(\eta|\bdy)$ denotes the distribution of $\eta$ conditionally
 to $\bdy$.\\

\begin{rem}
Even if we are working in a Bayesian framework, we do not want to
estimate the distributions of our parameters. Knowing the distribution
of the template image and its possible deformations is not of great
interest from an image analysis point of view. Indeed,  people are more interested, in particular in the medical imaging community, in an atlas
which characterises the populations of shapes that they
consider rather than its distribution. Moreover, the distribution of the
deformation law makes even less sense. This is the reason why we focus
on the MAP.\\
\end{rem}

In practice, to reach this estimator, we maximise this posterior
distribution using
a Stochastic Approximation EM (SAEM) algorithm coupled with a Monte
Carlo Markov Chain (MCMC) 
method. Indeed, due to  the intractable computation of the E step of
the EM algorithm introduced by \cite{DLR} 
encountered  in this complex non linear setting, we follow a
stochastic variation called SAEM proposed in \cite{DLM}. However,
again due to the expression of our model, the simulation required in
this algorithm cannot be performed directly. Therefore, we propose to
use some MCMC 
methods to reach this simulation as proposed by Kuhn and Lavielle in
\cite{kuhnlavielle}  and done for the one component
model in  \cite{aktdefmod}.
 Unfortunately, the direct generalisation  of the algorithm presented
 in \cite{aktdefmod} paper turns
out to be of no use in practice because of some trapping state
problems (cf. Subsection 3.2.). This suggests to go back to
some other extension of the SAEM  procedure.

\subsection{The SAEM algorithm using MCMC methods}

Let us first recall  the SAEM algorithm. It generates a sequence of
estimated parameters $(\e_k)_k$ which converges towards a critical
point of $\e \mapsto \log q(\bdy,\e)$ under some mild
assumptions (cf. \cite{DLM}). These critical points coincide with the
critical points of $\e \mapsto \log q_B(\e|\bdy)$.
The $k^{th}$ iteration  consists in three steps:
\begin{description}
\item[Simulation step~:] the
missing data, here the deformation parameters and the labels,
$(\bdbeta,\bdtau)$,
are drawn with respect to the distribution of $(\bdbeta, \bdtau)$ conditionally
to $\bdy$ denoted by
$\lstat_{\e}$, using the current parameter $\eta_{k-1}$
\begin{equation}
  \label{eq:simbeta}
  (\bdbeta_k,\bdtau_k) \sim\lstat_{\e_{k-1}}\,,
\end{equation}

\item[Stochastic approximation step~:] given $(\Delta_k)_k$ a decreasing
  sequence of positive step-sizes,
 a stochastic approximation is done on the quantity $\log
q(\bdy,\bdbeta,\bdtau,\eta)$,
using the simulated value of the missing data~:
\begin{equation}
\label{eq:appsto}
Q_{k}(\eta)=Q_{k-1}(\eta)+\Delta_{k}[\log
q(\bdy,\bdbeta_k,\bdtau_k,\eta)-Q_{k-1}(\eta)]\,,
\end{equation}

\item[Maximisation step~:] the parameters are updated  in the M-step,

  \begin{equation}
    \label{eq:updateeta}
    \eta_{k}=\Argmax\limits_\eta Q_{k}(\eta)\,.
  \end{equation}

\end{description}
Initial values  $Q_0$ and $\eta_0$ are arbitrarily chosen.

We notice that the density function of the model proposed in
paragraphs \ref{model} and \ref{bayes} belongs
 to the curved exponential family. That is to say that the
complete likelihood can be written as:
 $ q(\bdy,\bdbeta,\la, \e ) = \exp \left[ -\psi(\e) + \langle \tS(\bdbeta,\la),
  \phi(\e)\rangle \right] \,,$
where the sufficient statistic $\tS$ is a Borel function on $\R^N
� \T$ taking its
values in an open subset $\mathcal{S}$ of $\R^m$ and $\psi$, $\phi$
two Borel functions on $\Te � \varrho$. (Note that $\tS$, $\phi$ and $\psi$ may
depend also on $\bdy$, but since $\bdy$ will stay fixed in the sequel, we
omit this dependency).
Thanks to this property of our model, it is equivalent to do the stochastic
approximation on  the complete log-likelihood as well as on
the sufficient statistics. This yields equation \eqref{eq:appsto} to
be replaced by the following stochastic
approximation $s$ of the sufficient statistics $S$~:
\begin{equation}
\label{eq:appstoens}
s_k  = s_{k-1} + \Delta_{k} ( \tS(\bdbeta_k,\la_k) -  s_{k-1})�\, .
\end{equation}

We now introduce the following
function: $L : \mathcal{S} � \Te � \varrho \to \R $ as
$L(s ; \e)=  -\psi(\e) + \langle s,  \phi(\e)\rangle \ . \label{eq:a13}$
It has been proved in \cite{AAT} that there exists a critical function $
\hat{\e} : \mathcal{S} \to \Te � \varrho$ which is a zero of $\nabla L$. It is
straightforward to prove that this
function satisfies:
$  \label{eq:thetahat}
\forall \e \in \Te � \varrho,
\forall s \in \mathcal{S},
L(s;\hat{\e}(s)) \geq L(s; \e) $
so that the maximisation step \eqref{eq:updateeta} becomes:
$$\e_{k}=\hat{\e}(s_k)\,.$$

Concerning the simulation step, in our model, the simulation of the
missing variables with respect to the conditional distribution
$\lstat_\e$ cannot be carried out. Indeed, its probability density
function (pdf) has a close form but rather complicated ; it does not
correspond to some usual pdf.
One solution proposed in \cite{kuhnlavielle} for such cases is to
couple the  SAEM algorithm with Monte Carlo Markov Chain (MCMC)
method. However, we do not fit exactly into their requirements since
the missing variable $\bdbeta$ does not have a compact support.
We introduce an ergodic Markov chain whose stationary distribution is
the conditional distribution $ \lstat_\e$. We denote its transition
kernel  by $\ntrans_\e$. The simulation step \eqref{eq:simbeta} is
thus replaced by the following step:

\begin{equation}
  \label{eq:simbetaMCMC}
  (\bdbeta_k,\bdtau_k) \sim\ntrans_{\e_{k-1}}( (\bdbeta_{k-1},\bdtau_{k-1}),.)\,.
\end{equation}

The most common choice of kernel is an accept-reject step which
is carried out through a Metropolis-Hastings
algorithm. Unfortunately, in our particular setting, we deal with
large dimensions for the missing variables. This made us move to some
other kind of MCMC methods, like
a Gibbs sampler,  to simulate our missing variables.

\subsection{The transition of the MCMC method: a hybrid Gibbs sampler}
 If we consider
the full vector $(\bdbeta,
\bdtau)$ as a single vector of missing data, we can use the
hybrid Gibbs sampler on $\R^{N}� \T$ as follows.
 For any $b\in\mathbb{R}$ and $1\leq j\leq N$, let us denote
by $\betajb$ the unique configuration which is equal to $\bdbeta$ everywhere
 except the coordinate $j$ where $\betajb^j=b$ and by $\bdbetamj$ the
 vector $\bdbeta$
without the coordinate $j$.
Each coordinate of the deformation field $\bdbeta^j $ is updated using
a
Metropolis-Hastings step where the proposal is given by the conditional
distribution of $\bdbeta^j | \bdbetamj, \bdtau$ coming from the
current Gaussian
distribution with the corresponding parameters (pointed by $\bdtau$).
Then, the last coordinates corresponding to the missing variable
$\bdtau$ are drawn with respect to $ q(\bdtau|\bdbeta,\bdy,\e)$.

Even if this procedure provides an estimated
parameter sequence which would theoretically converge toward the MAP
estimator,
 in practice, as mentioned in \cite{Robert_MCMC}, it would take a
 quite long time to reach its limit because of the trapping state
problem:
when a small number of observations are assigned to a component, the
estimation of the component parameters is hardly concentrated and the
probability of changing the label of an image to this component or
from this component to another is really small (most of the time
under the computer precision).

We can interpret this from an image analysis viewpoint:
the first iteration of the algorithm gives a random label to the
training set and computes the corresponding maximiser $\eta = (\theta , \rho )$.
Then, for each image, according to its current label, it simulates a
deformation
field which only takes into account the parameters of this given component.
Indeed, the simulation of $\bdbeta$ through the Gibbs sampler involves a
proposal whose corresponding Markov chain has
$q(\bdbeta|\bdtau,\bdy,\eta)$ as stationary
distribution. Therefore, the deformation tries to match $\bdy$ to the
deformed template
of the given component $\bdtau$.
 The deformation field tries to get a better
connection between the component parameters and the observation, and
there is only small probability that the observation given
\textit{this} deformation field will be closer to another
component. The update of the label $\bdtau$ is therefore conditional
to this deformation which would not leave much chance to switch
component.

To overcome the trapping state problem, we will simulate the
optimal label, using as many Markov chains
in $\bdbeta$ as the
number of components so that each component has a corresponding
deformation which ``computes'' its distance to the observation. Then
we can simulate the optimal deformation corresponding to that optimal
label.

Since we aim to simulate $(\bdbeta,\bdtau)$ through a transition
kernel that has
$\qpost(\bdbeta,\bdtau|\bdy,\eta)$ as stationary distribution, we simulate
 $\bdtau$ with a kernel whose stationary distribution is
 $\qpost(\bdtau|\bdy,\eta)$ and then
 $\bdbeta$ through a transition kernel that has
$\qpost(\bdbeta |\bdtau,\bdy,\eta)$ as stationary distribution.

For the first step, we need to compute the weights
$q(t|y_i,\eta)\propto q(t,y_i|\eta)$ for all $1\leq t \leq \tau_m$ and all $1\leq  i \leq n $
which cannot be
easily  reached. However,
for any density function $\densMC$, for any image $y_i $ and for any
$1\leq t \leq \tau_m$, we have
\begin{equation}
  \label{eq:qgammay}
  q(t,y_i|\eta)= \left( \mathbb{E}_{\qpost(\beta|y_i,t,\eta)} \left[
\frac{\densMC(\beta )}{q(y_i,\beta,t|\eta)}
    \right] \right)^{-1} \, .
\end{equation}
Obviously the computation of this expectation w.r.t. the posterior
distribution is not tractable either
but we can approximate it by a Monte Carlo sum. However, we cannot
easily simulate  variables through
the posterior distribution $\qpost( \cdot |y_i,t,\e)$ as well, so we use some
realisations of an ergodic Markov chain having
$\qpost( \cdot |y_i,t,\e)$  as stationary distribution instead of some
independent realisations of this distribution.

The solution we propose is the following:
suppose we are at the $k^{th}$ iteration of the algorithm and let
$\e$ be the current parameters.
Given any initial deformation field $\xi_0\in$
$\R^{2k_g}$, we run,
for each component $t$,  the
hybrid Gibbs sampler $\ntrans_{\e,t}$ on $\R^{2k_g}$ $\nmc$ times so
that we get $\nmc$
elements $\xi_{t,i}=(\xi^{(l)} _{t,i})_{1\leq l\leq \nmc}$   of an
ergodic homogeneous Markov chain
whose stationary distribution is 
$q(\cdot |y_i,t, \e)$. Let us denote by $\xi_i=(\xi_{t,i})_{1\leq t\leq \tau_m}$ the
matrix of all the auxiliary variables. 
We then use these elements for the
computation of the weights  $p_{\nmc}(t|\xi_{i},y_i,\e)$  through a Monte Carlo sum:
\begin{equation}
  \label{eq:MCsum}
 p_{\nmc}(t|\xi_{i},y_i,\e)  \propto  \left( \frac{1}{\nmc} \sum\limits_{l=1}^\nmc
 \left[
   \frac{\densMC(\xi^{(l)} _{t,i})}{q(y_i,\xi^{(l)} _{t,i},t|\e)}
 \right]
 \right)^{-1} \,,
\end{equation}
where the normalisation is done such that their sum over $t$ equals
one, involving the dependence on all the auxiliary variables $\xi_i$.
The ergodic theorem ensures the
convergence of our approximation toward the expected value.
We then simulate $\bdtau$ through $\otimes_{i=1}^n  \sum\limits_{t=1}^{\tau_m}
p_{\nmc}(t|\xi_{i},y_i,\e) \delta_t$.

Concerning the second step,  we update $\bdbeta$ by re-running $\nmc$
times
the hybrid Gibbs sampler
$\ntrans_{\e,\la}$ on $\R^{N}$
starting from a random initial point $\bdbeta_0$ in a compact subset of $\R^N
$.  The size of $\nmc$ will depend on the iteration $k$ of the SAEM
algorithm  in a sense
that will be precised later, thus we now index it by $k$.

The density function $\densMC$ involved in the Monte Carlo sum above
needs to be 
specified to get the convergence result proved in the last section of
this paper. We show that using the prior on the deformation field
enables to get the sufficient conditions for convergence. This density
is the Gaussian density function and depends on the component we are
working with:
\begin{equation}
 \label{eq:densityf}
  \densMC_t (\xi) = \frac{1}{\sqrt{2\pi}^{2k_g}\sqrt{|\Gamma_{g,t}|}}
  \exp\left(-\frac{1}{2}\xi^t  \Gamma_{g,t}^{-1} \xi\right) \,.
\end{equation}

Algorithm \ref{Algo3}  shows the detailed iteration.

\begin{rem}
  The use of one simulation of $\bdbeta$ per component is a point that
  was already used in \cite{AAT} while computing the
  best matching, $\beta^*$, for all components by minimising the corresponding
  energies. This gives as many $\beta^*$ as components for each
  image. Then, according to these best matchings, it computed the
  best component which therefore pointed the matching to consider.
\end{rem}
\subsection{Truncation on random boundaries}
Since our missing data have a non-compact support, some of the
convergence assumptions
 of such algorithms \cite{kuhnlavielle} are not
satisfied. This leads to consider a truncation
algorithm as suggested in \cite{DLM} and extended in \cite{aktdefmod}.

Let
$(\Kapa_q)_{q\geq 0} $ be an increasing sequence of compact subsets of
$\mathcal{S}$ such as $\cup_{q\geq 0} \Kapa _q = \mathcal{S} $
and $ \Kapa _q \subset  \text{int}(\Kapa _{q+1}) , \forall q \geq 0$. Let
$\mathrm{K}$ be a compact subset of $\R^{N}$. Let $\ntrans_\e$ be a
transition kernel of an ergodic
 Markov chain on $\R^N$ having $\pi_\e$ as stationary distribution.
We construct the homogeneous Markov chain
 $((\bdbeta_k, \la_k, s_k, \kappa_k ))_{k\geq 0}$ as
explained in Algorithm
 \ref{AlgoMoulines}. As long as the stochastic approximation does not
wander out the current
compact set, we run our
"SAEM-MCMC" algorithm. As soon as this previous condition is
not satisfied, we reinitialise the sequences of $s$ and $(\bdbeta,
\bdtau)$  using
a projection (for more details see \cite{DLM}). The current compact is
then enlarge. To point toward the current compact, we use a counter sequence
$(\kappa_k)_k$ which remains unchanged when the previous condition  is
satisfied and increases to point toward a bigger compact when
re-projecting.

\begin{algorithm}
 \caption{Stochastic approximation with truncation on random boundaries}
\label{AlgoMoulines}
\begin{algorithmic}
\STATE Set $\bdbeta_0 \in \mathrm{K}$, $\bdtau_0 \in \T$, $s_0 \in \Kapa_0$ and $\kappa_0=0$.
\FORALL{$k\geq 1$}
\STATE\texttt{compute} $\bar{s} =
s_{k-1} + \Delta_{k} (
\tS(\bar{\bdbeta}, \bar{\bdtau}) -  s_{k-1}) $
\STATE \texttt{where} $(\bar{\bdbeta},\bar{\bdtau})$ \texttt{are sampled
from a transition kernel} $\ntrans_{\e_{k-1}}$ \texttt{(see Algorithm
\ref{Algo3})}.
\IF{ $\bar{s} \in \Kapa _{\kappa_{k-1}} $}
\STATE \texttt{set} $(s_k, \bdbeta_k,\bdtau_k) =
(\bar{s},\bar{\bdbeta},\bar{\bdtau})$
  \texttt{and}
$\kappa_k=\kappa_{k-1} $
\ELSE
\STATE \texttt{set}  $(s_k, \bdbeta_k,\bdtau_k) =
(\tilde{s},\tilde{\bdbeta},\bar{\bdtau})
 \in \Kapa_0
� \mathrm{K} � \T$ \texttt{and}
  $\kappa_k=\kappa_{k-1}+1 $
\STATE \texttt{and}
$(\tilde{s},\tilde{\bdbeta})$  \texttt{can be chosen through
  different ways (cf. \cite{DLM})}.
\ENDIF
\STATE $\eta_k=\Argmax\limits_\eta \hat\e(s_k)$.
\ENDFOR
\end{algorithmic}
\end{algorithm}
\begin{algorithm}
 \caption{Transition step $k\to k+1$ using a hybrid Gibbs sampler on $(\bdbeta,\la)$}
\label{Algo3}
\begin{algorithmic}
\REQUIRE $\e =\e_k$ , $\nmc=\nmck$
\FORALL{ $i = 1: n$}
\FORALL{ $t=1:\tau_m$ }
\STATE $\xi_{t,i}^{(0)} = \xi_{0}$
\FORALL{  $l=1:\nmc$}
\STATE $\xi = \xi_{t,i}^{(l-1)}$
\STATE \texttt{Gibbs sampler $\ntrans_{\e,t}$:}
\FORALL{ $j = 1: 2k_g$}
\STATE \texttt{Metropolis-Hastings procedure:}
\STATE $b \sim q(b|\xi^{-j},t,\e)$
\STATE Compute $
r_{j} (\xi^j,b;\xi^{-j},\eta,t) =\left[
  \frac{q(y_i|\xi_{b\to j},t,\eta)}{q(y_i|\xi,t,\eta)}\land
  1\right]
$
\STATE $u \sim \mathcal{U}[0,1]$
\IF {$u<r_{j} (\xi^j,b;\xi^{-j},\eta,t)$} 
\STATE  $\xi^j=b$
\ENDIF
\ENDFOR
\STATE $\xi_{t,i}^{(l)} = \xi$
\ENDFOR
\begin{equation*}
p_{\nmc}(t|\xi_{i},y_i,\e ) \propto \left( \frac{1}{\nmc} \sum\limits_{l=1}^{\nmc}
 \left[
   \frac{\densMC_t(\xi_{t,i} ^{(l)})}{q(y_i,\xi_{t,i}^{(l)},t|\e)}
 \right]
 \right)^{-1}
\end{equation*}
\ENDFOR
\ENDFOR
\STATE
\begin{equation*}
\la_{k+1} \sim \otimes_{i=1}^n  \sum\limits_{t=1}^{\tau_m}
p_{\nmc}(t|\xi_{i}, y_i,\e) \delta_t \hspace{0.3cm}\text{and}\hspace{0.3cm}
%
%
\bdbeta_{k+1} \sim \ntrans_{\e,\la_{k+1}}^{\nmc}(\bdbeta_{0}).
\end{equation*}
\end{algorithmic}
\end{algorithm}
\section{Convergence theorem of the multicomponent procedure}
\label{convergence}
In this particular section the variances of the components
$(\si^2_t)_{1\leq t\leq \tau_m}$ are fixed. Alleviating this condition is
not straightforward and is an issue of our future work.

To prove the convergence of our parameter estimate toward the MAP,
we have to go back to a convergence theorem which deals with general
stochastic approximations. Indeed, the SAEM-MCMC algorithm introduced
and detailed above is a Robbins-Monro type stochastic approximation
procedure. One common tool to prove the w.p.1 convergence of such a
stochastic approximation has been introduced by Kushner and Clark in
\cite{kushnerclark78}. However, some of the assumptions they require
are intractable with our procedure (in particular concerning the mean
field defined below). This leads us to slightly adapt the convergence
theorem for stochastic approximations given in \cite{DLM}.

We consider the following Robbins-Monro stochastic approximation
procedure:

\begin{equation}
\label{robbinsmonroe}
s_k  = s_{k-1} + \Delta_k( h(s_{k-1}) + e_k + r_k) \,,
\end{equation}
where $(e_k)_{k\geq 1}$ and  $(r_k)_{k\geq 1}$ are random processes
defined on the same probability space taking their values in an open
subset $\mathcal{S}$ of $\R^ {n_s}$; $h$ is referred to as the mean field
of the algorithm;
$(r_k)_{k\geq 1}$ is a remainder term and  $(e_k)_{k\geq 1}$ is a
stochastic excitation.

To be able to get a convergence result, we consider the
truncated sequence $(s_k)_k$ defined as follow: let $\mathcal{S}_a \subset
\mathcal{S}$ and  $\bar{s}_k  =
s_{k-1} + \Delta_k h(s_{k-1}) + \Delta_k e_k + \Delta_k r_k \,,$ where

\begin{equation}\label{sktruncated}
\begin{array}{l}
\text{ if } \bar{s}_k \in \Kapa_{\kappa_{k-1}}  \left\{
\begin{array}{l}
s_k= \bar{s}_k \,,\\
\kappa_k = \kappa_{k-1}\,,
\end{array}\right.\\
\text{ if } \bar{s}_k \notin  \Kapa_{\kappa_{k-1}}  \left\{
\begin{array}{l}
s_k= \tilde{s}_k \in \mathcal{S}_a\,,\\
\kappa_k = \kappa_{k-1}+1 \,.
\end{array}\right.
\end{array}
\end{equation}
The projection  $\tilde{s}_k $ can be made through
  different ways (cf. \cite{DLM}).

We will use
Delyon's theorem which gives sufficient conditions for the sequence
$(s_k)_{k\geq 0}$ truncated on random boundaries to converge with
probability one. The theorem we state here  is a
generalisation of the theorem presented in \cite{DLM}. Indeed, we have
add the existence of an absorbing set for the stochastic approximation
sequence. The proof of this theorem can be carried out the same way
Delyon et al. do theirs adding the absorbing set. This is why it is not
detailed here.
\begin{theorem} \label{SA}
We consider the sequence $(s_k)_{k\geq 0}$  given by the truncated
procedure \eqref{sktruncated}.
Assume that :
\begin{description}
\item[(SA0')] There exists a closed convex set $\mathcal{S}_a \subset
    \mathcal{S}$ such that for all $k\geq0$, $s_k \in \mathcal{S}_a$ w.p.1.
\item[(SA1)] $(\Delta_k)_{k\geq 1}$ is a decreasing sequence of
  positive numbers such that $\sum\limits_{k=1}^\infty \Delta_k = \infty$.
\item[(SA2)]The vector field $h$ is continuous on
  $\mathcal{S}$ ou $\mathcal{S}_a$ and there
  exists a continuously differentiable function $w: \mathcal{S} \to
  \R$ such that
\begin{description}
\item[(i)]for all $s \in \mathcal{S}, F(s) = \langle \partial_s
  w(s),h(s)\rangle \leq 0$.
\item[(ii)]$int(w(\mathcal{L}')) = \emptyset$, where
  $\mathcal{L}' \triangleq \{ s \in \mathcal{S} : F(s)=0\}$.
\end{description}
  \item[(STAB1')]  There exist a continuous differentiable function
$W~:
   \R^N \to \R$ and a compact set $\Kapa$ such that
 \begin{description}
\item[(i)] For all $c\geq 0$, we have $\mathcal{W}_c \cap \mathcal{S}_a$ is a compact subset of $\mathcal{S}$ where $\mathcal{W}_c=\{ s\in \mathcal{S} : \ W(s)\leq c \} $
  is a level set.
\item[(ii)] $\langle \partial_s W(s), h(s)\rangle <0 $, for all $s \in \mathcal{S}\setminus  \Kapa $.
\end{description}
 \item[(STAB2)] For any positive integer $M$, w.p.1  $\lim\limits_{p\to \infty}
   \sum\limits_{k=1}^p \Delta_k e_k\mathds{1}_{W(s_{k-1}) \leq M} $ exists and is
   finite and  w.p.1 $\limsup\limits_{k\to \infty} |r_k|\mathds{1}_{W(s_{k-1}) \leq M}  =0$.


\end{description}

Then,  w.p.1, $\limsup\limits_{k\to \infty} d(s_k,\mathcal{L}')=0$.
\end{theorem}

Assumption (\textbf{SA2}), which involves a Lyapunov function $w$,
replaces the usual condition that, w.p.$1$, the
sequence comes back infinitely often in a compact set which is not
easy to check in practice. In addition, assumptions (\textbf{STAB1'}) and
(\textbf{STAB2}) give a recurrent condition introducing a Lyapunov
function $W$ which controls the excursions outside the compact 
sets. 
The two Lyapunov functions $w$ and $W$ do not need to be the
same. Another interesting point is that the truncation does not change
the mean field and therefore the stationary points of the sequence. \\

This theorem does not ensure the convergence of the sequence to a
maximum of the likelihood but to one of its critical points. To ensure
that the critical point reached is a maximum, we would have to satisfy
two other conditions (called (\textbf{LOC1-2}) in \cite{DLM}) which
are typical conditions. That is to say, it requires that the critical
points are isolated and for every stationary points
$s^*\in\mathcal{L}'$ the Hessian matrix of the observed $\log$-likelihood
is negative definite. \\

We now want to apply this theorem to prove the convergence of our
``SAEM like'' procedure where
the missing
variables are not simulated through the posterior density function
but by a kernel which can be as close as wanted -increasing $\nmc_k$- to this
posterior law (generalising Theorem 3 in \cite{DLM}).

Let us consider the following stochastic approximation: $(\bdbeta_k,
\bdtau_k)$ are simulated
by the transition kernel described in the previous
section and
\begin{equation*}
s_{k}=  s_{k-1} + \Delta_k (\tS(\bdbeta_k, \bdtau_k) - s_{k-1}) \,,
\end{equation*}
which can be connected to the Robbins-Monro procedure using the
notations introduced in \cite{DLM}: let $\mathcal{F}=
 (\mathcal{F}_k)_{k\geq 1}$ be the filtration
 where $\mathcal{F}_k$ is the $\si-$algebra generated by the random
 variables $(S_0,   \bdbeta_1, \dots, \bdbeta_k,  \la_1, \dots,
 \la_k)$,  $\mathbb{E}_{\lstat_\e} $ is the expectation with respect
 to
the posterior distribution $\lstat_\e$ and

\begin{eqnarray*}
h(s_{k-1} )&=& \mathbb{E}_{\lstat_{\hat{\e}(s_{k-1})}} \left[ \tS (\bdbeta, \bdtau )
\right] -s_{k-1} \,,\\
e_{k}  & = & \tS ( \bdbeta_{k}, \bdtau_{k})
- \mathbb{E}
\left[ \tS( \bdbeta_k, \bdtau_k ) | \mathcal{F}_{k-1}
 \right]\,,\\
r_k&=& \mathbb{E}
\left[
  \tS   (\bdbeta_k, \bdtau_k)| \mathcal{F}_{k-1}
 \right] - \mathbb{E}_{\lstat_{\hat\e(s_{k-1})}} \left[ \tS (\bdbeta, \bdtau)
\right] \,.
\end{eqnarray*}
\begin{theorem}
\label{saemlike}
 Let $w(s)=-l (
\hat{\e}(s))$ where  $l(\eta) = \log \sum\limits_{\la} \int_{\R^N} q(\bdy,\bdbeta,\la,\eta)
d\bdbeta $
and $h(s)=\sum\limits_{\la} \int_{\R^N}
(S(\bdbeta,\bdtau )-s)\lstat_{\hat{\e}(s)}(\bdbeta,\la)d\bdbeta$ for
$s\in \mathcal{S}$.
Assume that:
\begin{description}
\item[(A1)] the sequences $(\Delta_k)_{k\geq 1} $ and $(J_k)_{k\geq
    1} $ satisfy~:
\begin{description}
\item[(i)]
  $(\Delta_k)_{k\geq 1} $ is non-increasing,
  positive,
$\sum\limits_{k=1}^\infty \Delta_k =\infty \text{ and }
 \sum\limits_{k=1}^\infty \Delta_k^2 <\infty\,.$
 \item[(ii)]
 $(J_k)_{k\geq 1} $ takes its values in the set of positive
integers and
 $\lim\limits_{k\to \infty} J_k = \infty\,.$
\end{description}
\vspace{0.3cm}
\item[(A2)]  $\mathcal{L}' \triangleq \left\{
s \in \mathcal{S} , \left\langle
\partial_s w(s), h(s) \right\rangle =0
\right\}$ is included in a level set of $w$.
\end{description}
 Let   $(s_k)_{k\geq 0}$ be the truncated sequence defined in equation
 (\ref{sktruncated}),
     $\mathrm{K}$ a compact set of $\R^N$ and $\Kapa_0\subset S(\mathbb{R}^N)$ a
     compact subset of $\mathcal{S}$.  Then, for
     all  $\bdbeta_0 \in \mathrm{K}$, $\la_0\in\T$ and
    $s_0\in \Kapa_0$, we have
$$\lim\limits_{k\to\infty}
    d(s_k,\mathcal{L}')=0\ \bar{\mathbb{P}}_{\bdbeta_0,\bdtau_0, s_0,0}
    \text{ -a.s.}\,,$$
    where $ \bar{\mathbb{P}}_{\bdbeta_0, \bdtau_0, s_0,0} $ is the
    probability measure
    associated with the chain $(Z_k=(\bdbeta_k,\bdtau_k,
    s_k,\kappa_k))_{k\geq 0}$ starting at $(\bdbeta_0,\bdtau_0,s_0,0) $.

\end{theorem}
\vspace{0.3cm}
The first assumption which concerns the two sequences involved in the
algorithm, is not restrictive at all since these sequences
can be chosen arbitrarily.

The second assumption, however, is more complex. This is required
to satisfy the assumptions of Theorem \ref{SA}. This is a condition we
have not proved yet and is part of our future work.

\vspace{0.3cm}
\begin{proof}
  The proof of this theorem is given in Section \ref{proof}. We give here a
  quick sketch to emphasise the main difficulties and differences
  between our proof and  the convergence proof of the SAEM algorithm given
  in \cite{DLM}.

Even if the only algorithmic
difference between our algorithm and the SAEM algorithm is the
simulation of the missing data which is not done with respect to the
posterior law $\qpost(\bdbeta,\bdtau|\bdy,\eta)$ but through an
approximation which can be arbitrarily close, this yields a very
different proof. Indeed, whereas for
the SAEM algorithm, the stochastic
approximation leads to a Robbins-Monro type equation
\eqref{robbinsmonroe} with no residual term
$r_k$, our method induces
one. The first difficulty is therefore to prove that this residual term
tends to $0$ while the number of iterations $k$ tends to infinity.
Our proof is decomposed into two part, the first one concerning the
deformation variable $\bdbeta$ and the second one the label $\bdtau$.
The first term requires to prove the geometric ergodicity of the Markov
chain in $\bdbeta$ generated through our kernel. For this purpose, we
prove some
typical sufficient conditions which include the existence of a small
set for the transition kernel and a drift condition.
Then, we use for the second term some concentration inequalities for
non stationary Markov chains to
prove that the kernel
associated with the label distribution converges toward the posterior
distribution $\qpost(\bdtau|\bdy, \eta)$.

The second difficulty is to prove the convergence of the
excitation term $e_k$. This can be carried out as in \cite{DLM} using the
properties of our Markov chain and some martingale limits properties.
\end{proof}

\begin{Cor}
  Under the assumptions of Theorem \ref{saemlike} we have for
     all  $\bdbeta_0 \in \mathrm{K}$, $\bdtau_0\in\T, s_0 \in \mathcal{S}_a$ and
    $\e_0\in \Te � \varrho $,
$$
\lim\limits_{k\to
  \infty} d(\e_k,\mathcal{L})=0 \ \ \bar{\mathbb{P}}_{\bdbeta_0,\bdtau_0, s_0,0}
\text{-a.s} \,,
$$
 where $ \bar{\mathbb{P}}_{\bdbeta_0, \bdtau_0, s_0,0,} $ is the
 probability measure
    associated with the chain $(Z_k=(\bdbeta_k,\bdtau_k,
    s_k,\kappa_k))_{k\geq 0}$ starting at
    $(\bdbeta_0,\bdtau_0,s_0,0) $ and $\mathcal{L}\triangleq
 \{�\e\in\hat{\e}(\mathcal{S}), �\frac{\partial l}{\partial
    \e}(\e)=0\}$.

\end{Cor}\vspace{0.3cm}
\begin{proof}
This is a direct consequence of the smoothness of the function $ s\mapsto
\hat\e(s)$ on $\mathcal{S}$ and of Lemma 2 of \cite{DLM} which links the
sets $\mathcal{L}$ and $\mathcal{L}'$.
\end{proof}

\section{Experiments}\label{experiments}

\subsection{USPS database}
\begin{figure}[htbp]
\centerline{\epsfxsize=7cm \epsfbox{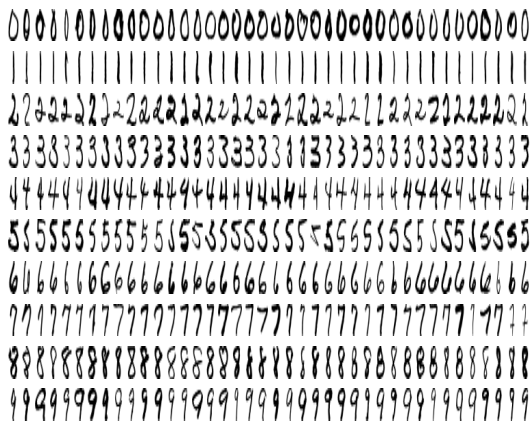}}
\caption{Some examples of the training set: 40 images per class
  (inverse video). } \label{fig-training40}
\end{figure}

To illustrate the previous algorithm for the deformable template
model, we are considering handwritten digit images. For each digit,
referred as class later, we learn  two templates, the
corresponding noise variances and the geometric covariance matrices. We
use the USPS
database which contains a training set of around 7000 images.
Each picture is a $(16� 16)$ gray level image with intensity in
$[0,2]$ where $0$ corresponds to the black background. In Figure \ref{fig-training40} we show some of the training images used for the statistical estimation.


\subsubsection{General setting of the algorithm}
 A natural choice for the prior laws on $\alpha$ and $\Gamma_g$ is to set
 $0$ for the mean on $\alpha$ and to induce the two covariance matrices
 by the metric of the spaces
 $V_p$ and $V_g$ involving the correlation between the landmarks
 through the kernels:
 define the square matrices
 $M_p(k,k')=K_p(\pk,\pkp) \ \forall 1 \leq k,k'\leq k_p\,,$ and $
 M_g(k,k')=K_g(\gk,\gkp) \ \forall 1 \leq k,k'\leq k_g\,.$
 Then $\Gammap = M_p^{-1}$ and  $\Gammag = M_g^{-1}$.
 In our experiments, we have chosen Gaussian kernels for both $K_p$ and
 $K_g$, where the standard deviations are fixed: $\sigma_p = 0.2$ and
  $\sigma_g=0.12$ (for an estimation on $[-1.5,1.5]^2$ and $[-1,1]$ respectively).

 For the stochastic approximation step-size, we allow a heating period
 $k_h$ 
 which corresponds to the absence of memory for the first
 iterations. This allows the Markov chain to reach an area of interest
 in the posterior probability density function $q(\bdbeta,\la|\bdy,
 \eta)$ 
before
 exploring this particular region.
In the experiments presented, the heating time $k_h$ lasts up to
 $150$ iterations and the whole algorithm is stopped at, at most, $200$
 iterations depending on the data set (noisy or not). This number of
 iterations corresponds to a point when the
 convergence is reached. We choose, as suggested in \cite{DLM} 
the step-size sequence as $\Delta_k
 = 1$ for all $1\leq k \leq k_h$ and $\Delta_k 
 = (k-k_h)^{-0.6}$ otherwise.

The multicomponent case has to face the problem of its computational
time. Indeed, as we have to approximate the posterior density by
running $\nmc_k$ elements of $\tau_m$ independent Markov chains, the
computation time increases linearly with $\nmc_k$. In our experiments,
we have chosen a fixed $\nmc$ for every EM iteration, $\nmc=50$. This
is enough to get a good approximation thanks to the coupling between
the iterations of the SAEM algorithm and the iterations of the Markov
chains. Indeed, even if the parameter $\eta$ is modified along the SAEM
iterations, its successive jumps are small enough to ensure the
convergence of the MCMC method. Intuitively speaking, it is equivalent
to consider not only $50$
iterations of the MCMC method but $50$ times the number of SAEM
iterations.

 \subsubsection{The estimated templates}
 We are showing here the results of the statistical learning algorithm
  for our generative model.
 To
 avoid the problems shown in \cite{aktdefmod}, we choose
 the same initialisation of the template parameter $\alpha$ as they did,
 that is to say,
we set
 the initial value of $\alpha$
 such that the corresponding $I_\alpha$ is the mean of the gray-level
 training images.

\begin{figure}[htbp]
  \centerline{\epsfxsize=3cm
    \epsfbox{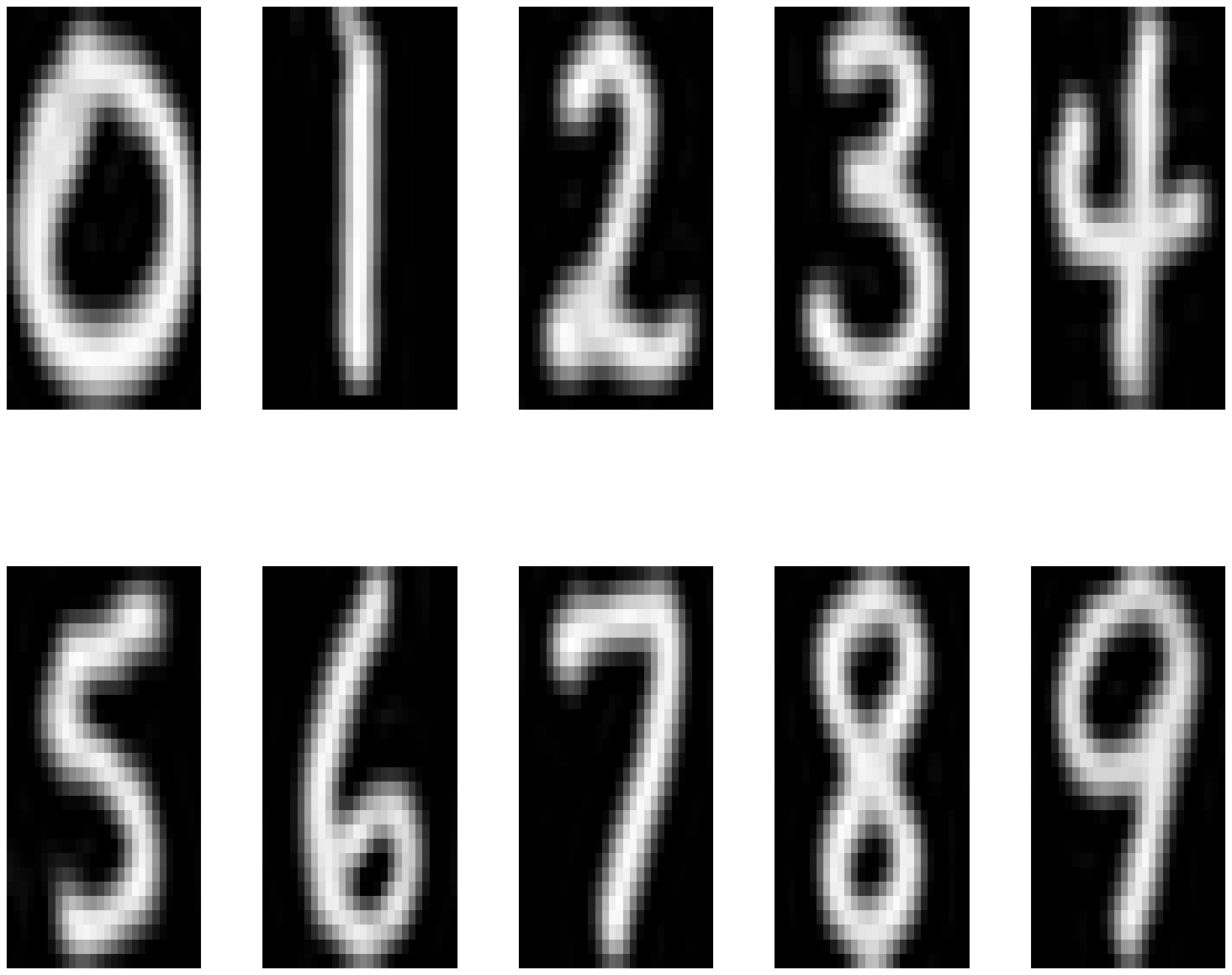} \hspace{0.5cm} \epsfxsize=3cm
    \epsfbox{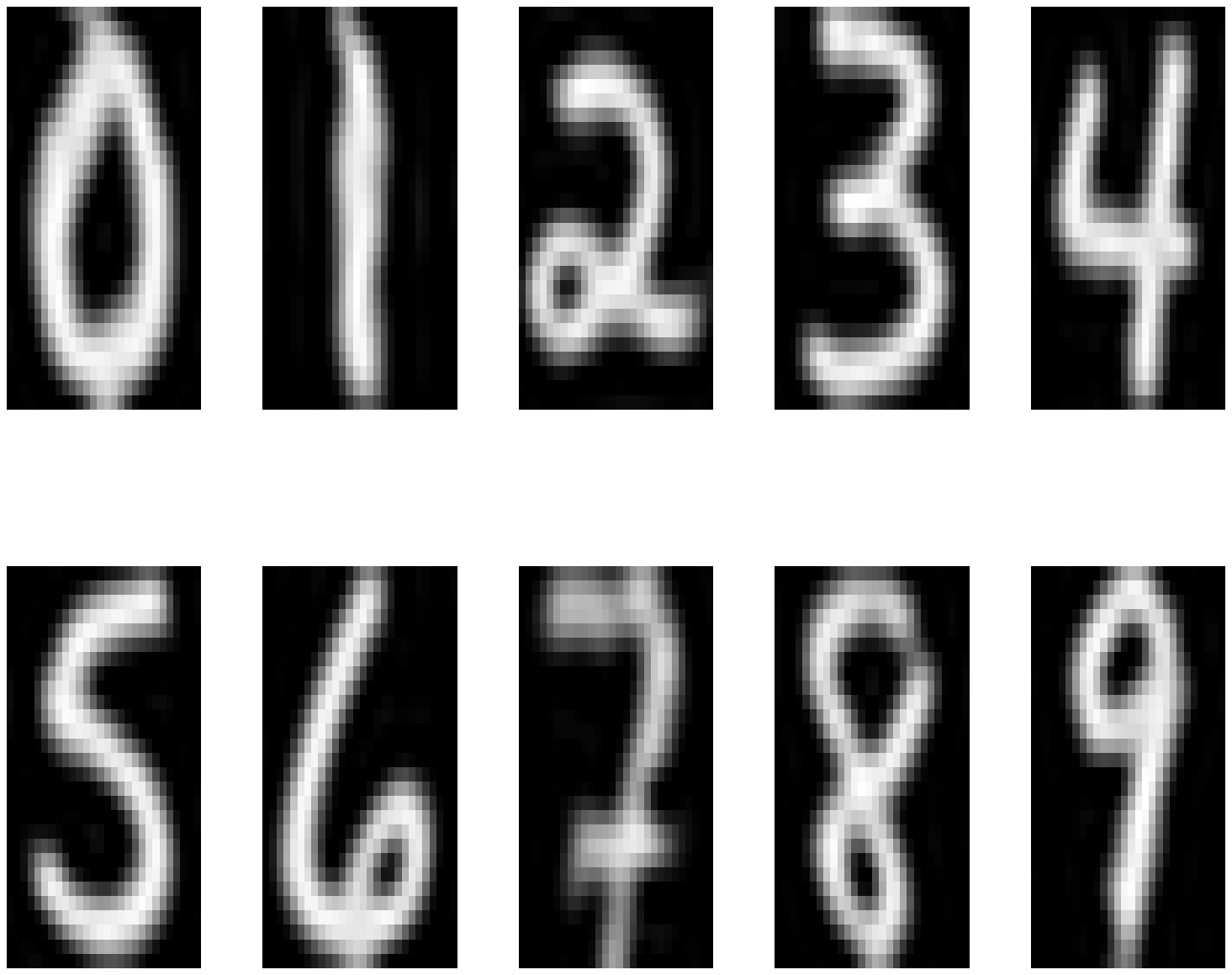}
}
\caption{Estimated prototypes of the two components model for each digit (40
  images per class; 100 iterations; two components per class).}
\label{fig:templatemulti1run}
\end{figure}

\begin{figure}[htbp]
  \centerline{\epsfxsize=3cm
    \epsfbox{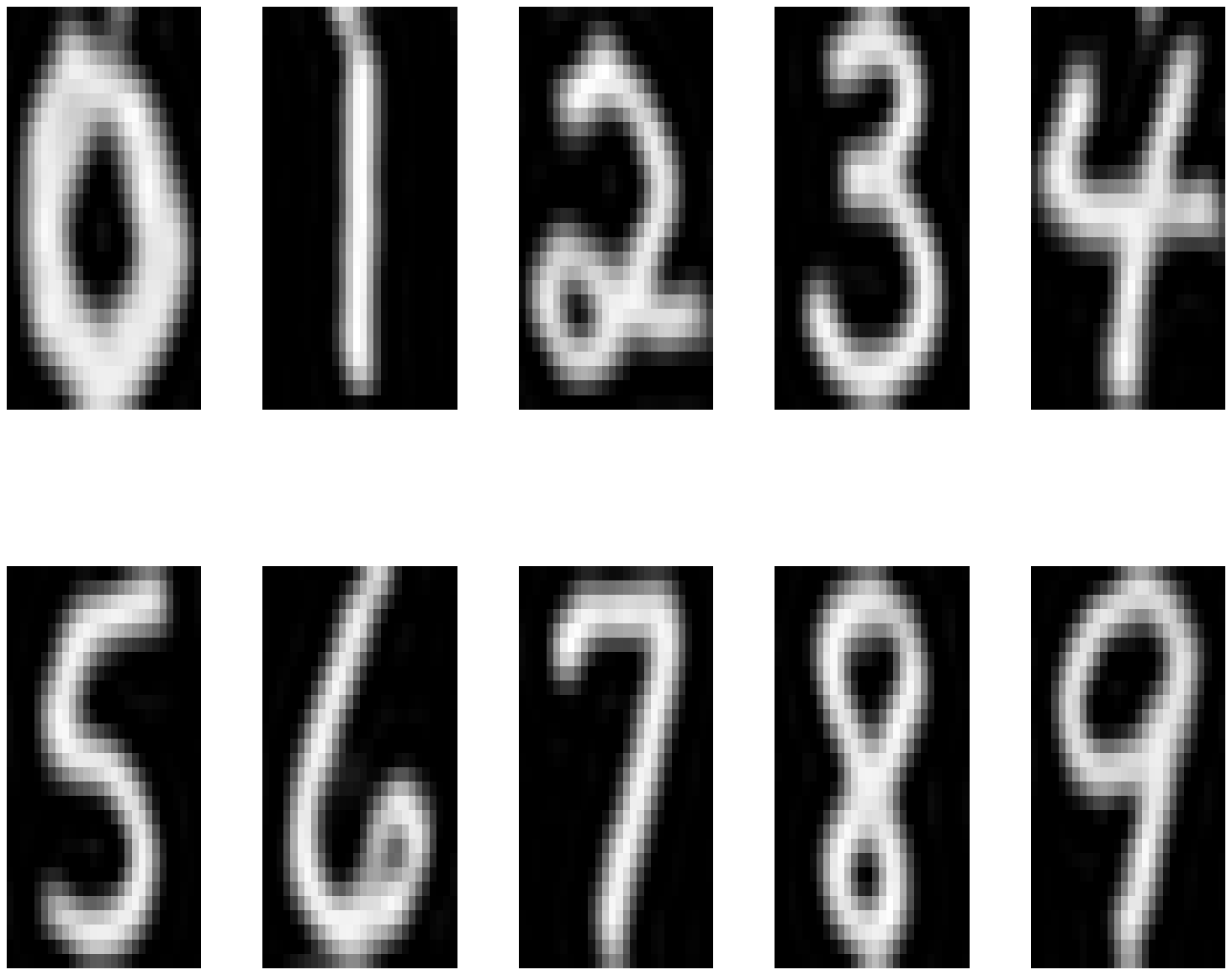} \hspace{0.5cm} \epsfxsize=3cm
    \epsfbox{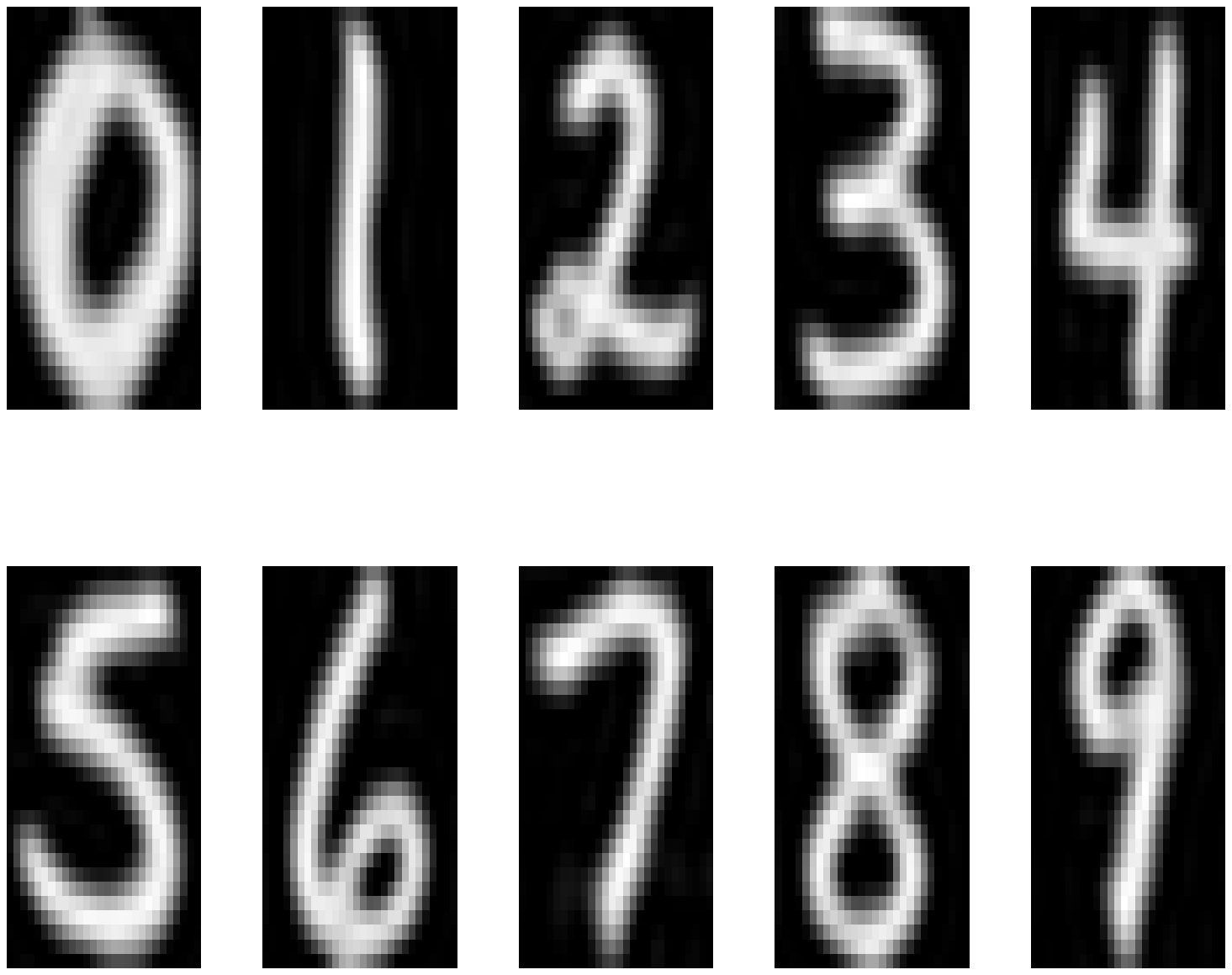}
}
\caption{Estimated prototypes of the two components model for each digit (40
  images per class, second random sample).}
\label{fig:templatemulti2run}
\end{figure}

In Figure \ref{fig:templatemulti1run}, we show the two estimated templates
obtained by the multicomponent procedure with $40$
training examples per class.
It appears that, as for the mode
approximation algorithm which results are presented on this database
in \cite{AAT}, the two components reached are
meaningful, such as the 2 with and without loop or American and
European 7. They even look alike.

In Figure \ref{fig:templatemulti2run}, we show a second run of the
algorithm with a different database, the training images are randomly
selected in the whole USPS training set. We can see that there are
some variability, in particular for digit $7$ where there were no
European $7$ in this training set. This generates two different clusters
still relevant for this digit. The other digits are quite stable, in
particular the strongly constrained ones (like $3$, $5$, $8$ or $9$).

\subsubsection{The photometric noise variance}

\begin{center}
\begin{figure}[htbp]

 \includegraphics[width=7.8cm,height=6cm]{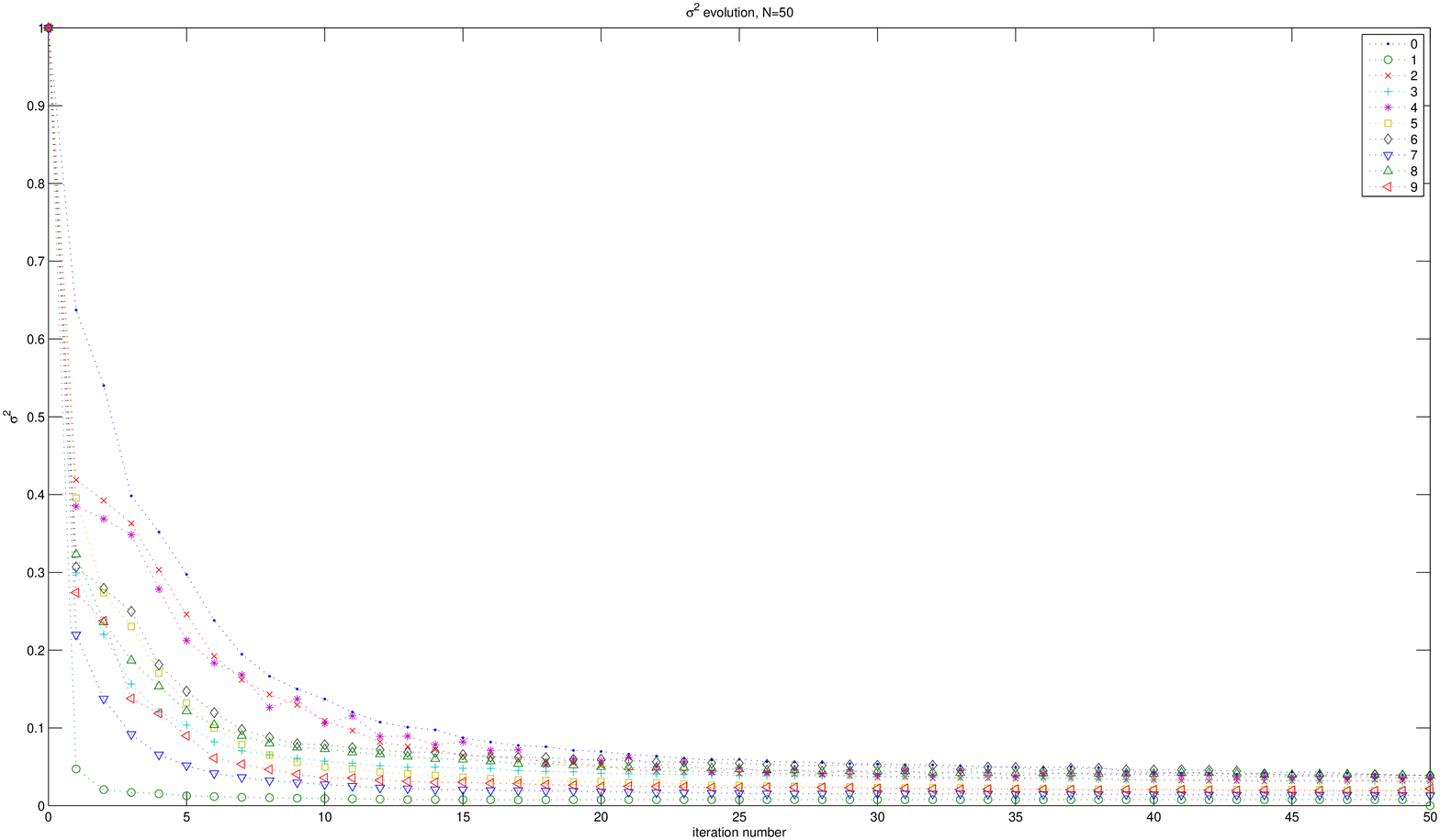} \hspace{-0.9cm}
 \includegraphics[width=7.8cm,height=6cm]{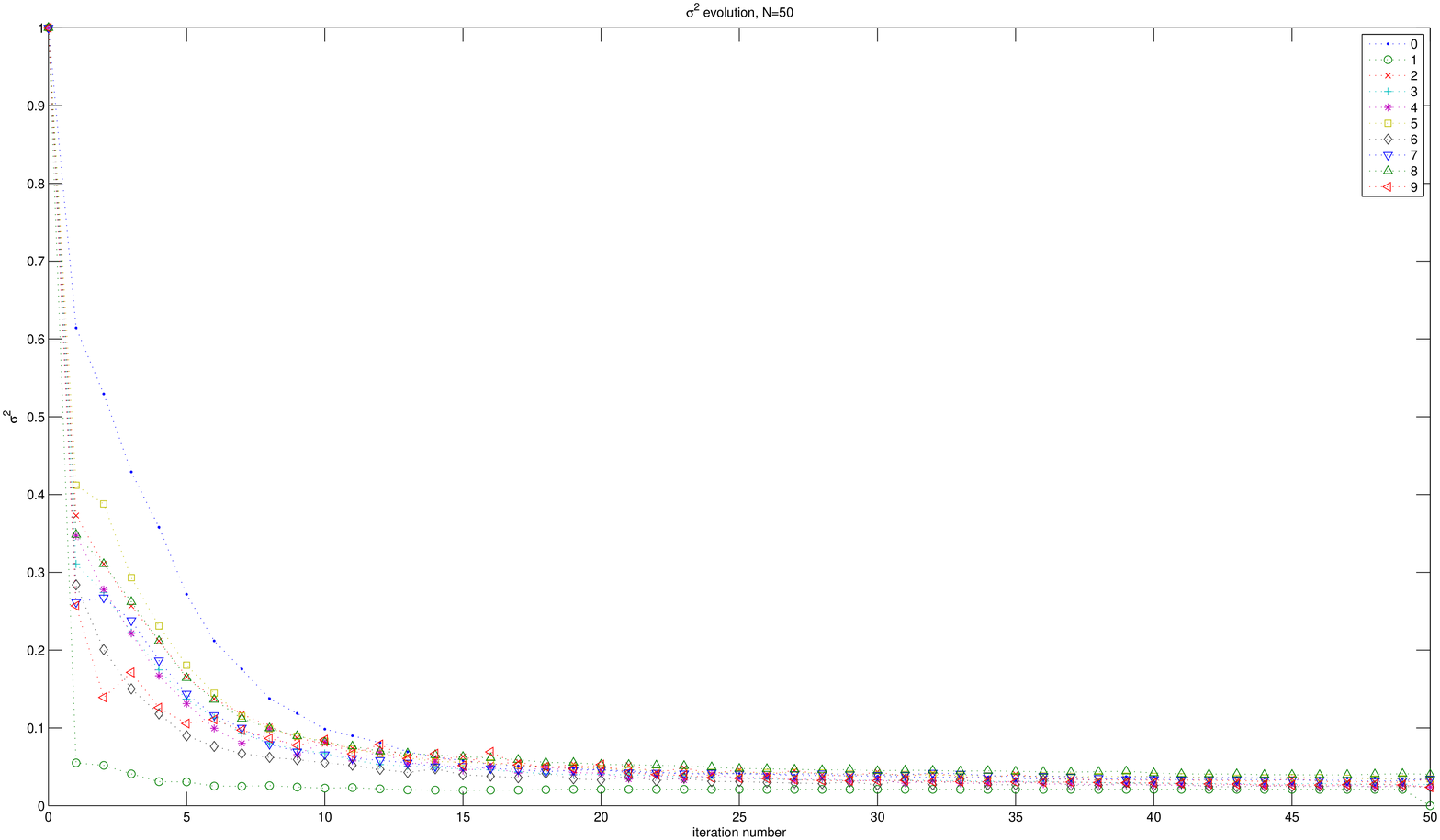}
\caption{Evolution of the two cluster variances along the iterations.}
\label{fig:evolsig}
\end{figure}
\end{center}

Even if we prove the convergence result for  fixed component noise
variances, we still try to learn them in the experiments.
 The same behaviour for our stochastic EM as for the mode approximation
 EM algorithm done in \cite{AAT} is observed for the noise
 variances. Indeed,
allowing the decomposition of the class into components enables the model
to better fit  the data yielding a lower residual noise. In
addition, the stochastic algorithm enables to look around the whole
posterior distribution and not only focusing on its mode which
increases the accuracy of the geometric covariance and the template
estimation. This yields lower noise required to explain the gap
between the model and the truth. The evolution of the estimated
variances for the two components of each digits are presented in
Figure \ref{fig:evolsig}.

The convergence of this variance for some
very constrained digits like digit $1$ is faster. This is due to the well
defined templates and geometric variability in the class which can be
easily captured. Therefore, a
very low level of noise is required very quickly.
On the other hand, some very variable digits like digit $2$ are slower
to converge. The huge geometric variability adding to very complex
shapes for the templates lead to a more difficult estimation and
therefore more iterations before convergence.

Last point that can be noticed is the convergence of the European $7$
which looks slower than the other component (American $7$). The reason
of this behaviour is that there are only two images of such a $7$ in
the training set and
it takes a longer time for the algorithm to put together and only
together these two shapes so that the clustering is better with
respect to the likelihood. The other $7$ does not suffer from this
problem and converges faster.

\subsubsection{The estimated geometric distribution}
\begin{center}
\begin{figure}[htbp]
\begin{minipage}[h]{19cm}
 \begin{tabular}[h]{cc}
 \includegraphics[width=16cm,height=1cm]{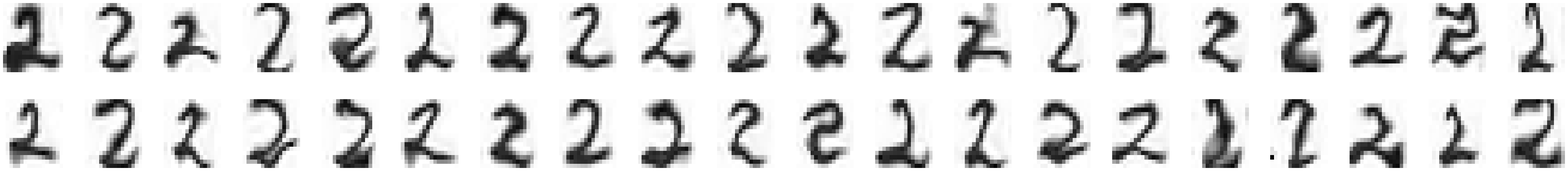}&
  \\
 \includegraphics[width=16cm,height=1cm]{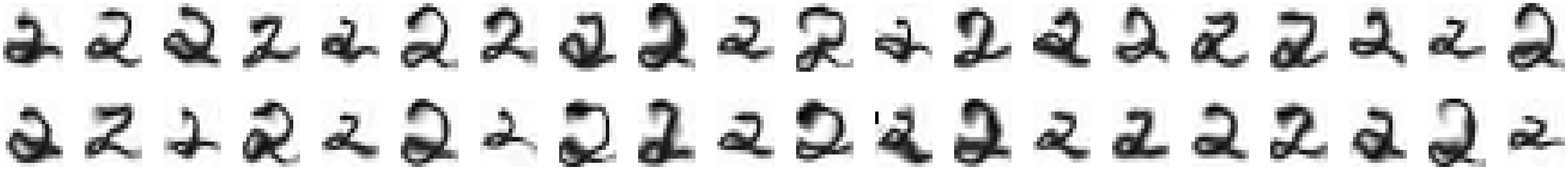}& \\

 \includegraphics[width=16cm,height=1cm]{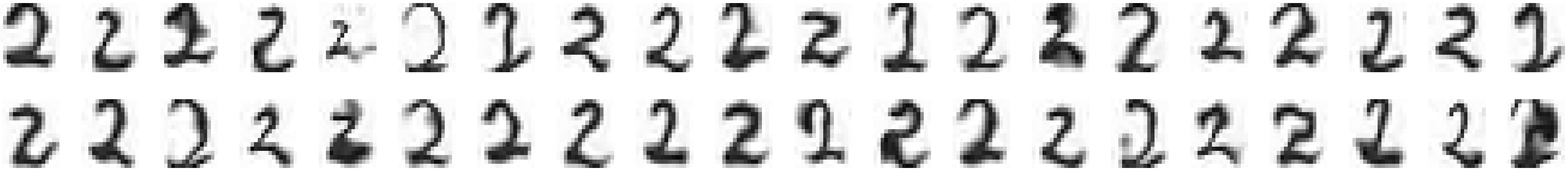}&
\end{tabular}
\end{minipage}
\caption{Some synthetic examples of the components of digit 2: First
  four rows: templates of the two components deformed through some
  deformation field $\beta$ and $-\beta$ drawn
  from their respective geometric covariance. Two last row:
  template of the first component from Figure
  \ref{fig:templatemulti1run} with
  deformations drawn with respect to the second component covariance matrix.}
\label{fig:samplemulti}
\end{figure}
\end{center}

To be able to compare the learnt geometry, we draw some
synthetic examples using the mixture model with the learnt
parameters. Even when the templates look similar, the separation
between two components can be justified by the different geometry
distributions. To show the effects of the geometry on the components,
we have drawn some ``2''
with their respective parameters in the four top rows of Figure
\ref{fig:samplemulti}.

For each component, we have drawn the
deformation given by the variable $\beta$ and its opposite $-\beta$ since, as
soon as one is learnt, because of the symmetry of the centred Gaussian
distribution, the opposite deformation is learnt at the same
time. This is why sometimes, one of the two looks strange whereas the
other looks like some element of the training set.

The simulation is done using a common standard Gaussian distribution
which is then multiplied by a square root of the covariance matrix we
want to apply.
We
can see the effects of the covariance matrix on both templates and the
large variability learnt. This
has to be compared with the bottom rows of Figure \ref{fig:samplemulti}, where
the two samples are drawn on the one  template but with
the
covariance matrix of the other one.
Even if these six
lines represent some ``2''s, the bottom ones suffer from the geometrical
tendency of the other cluster and do not look as natural. This shows
the variability of the models into classes.


\subsection{Medical images}

\begin{center}
\begin{figure}[tb]
\hspace{-1cm}  \begin{minipage}[h]{1.0\linewidth}
    \begin{tabular}[h]{c}
      \includegraphics[width=13cm]{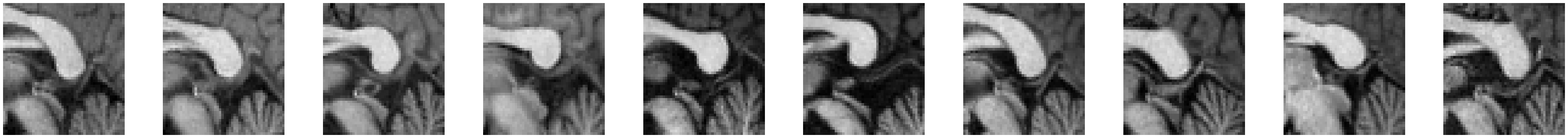}\\
      \begin{tabular}[h]{ccccc}
        \begin{tabular}[h]{c}
          \includegraphics[width=2.5cm]{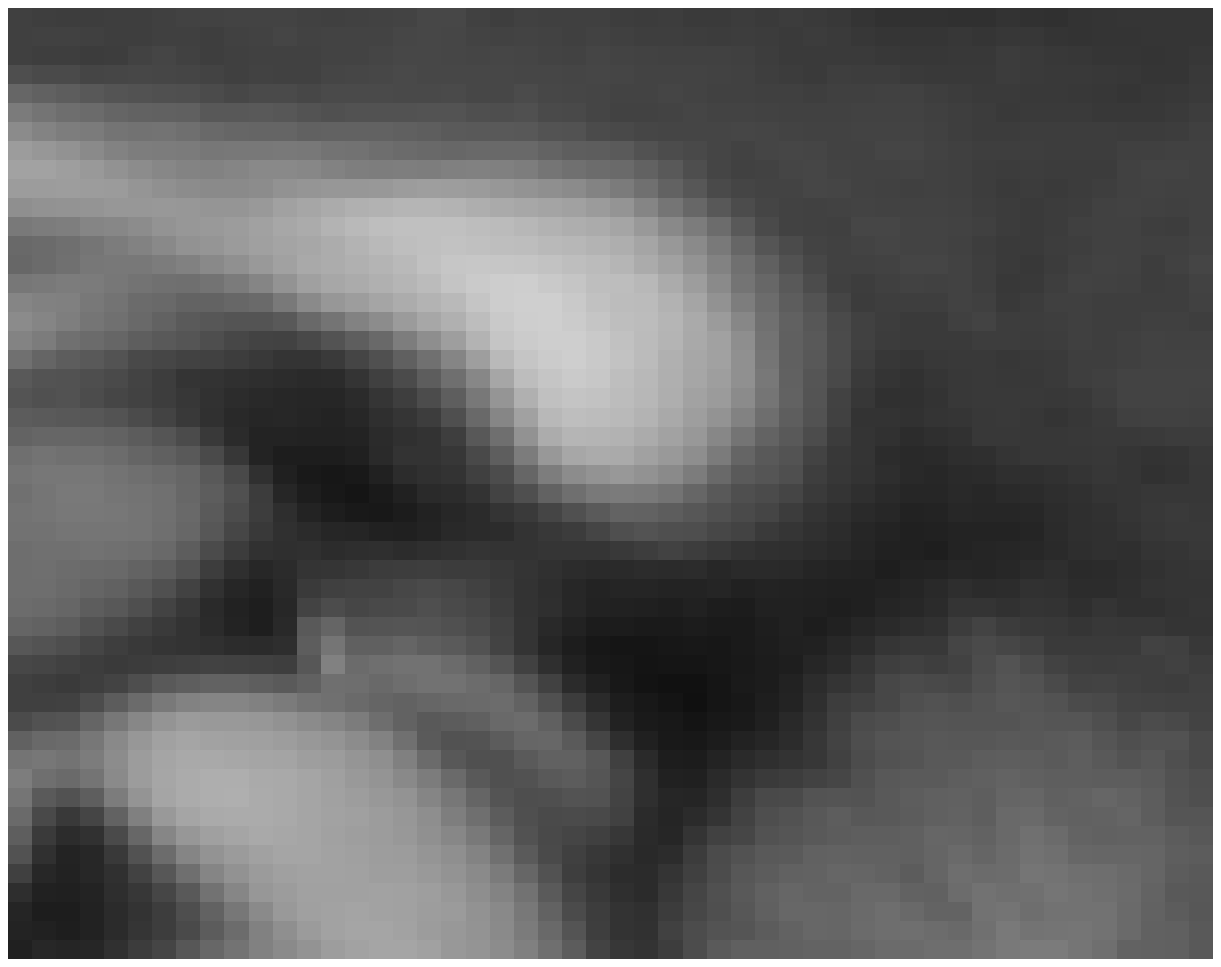}\\
          (a)
        \end{tabular}
        &\hspace{-0.8cm}
        \begin{tabular}[h]{c}
          \includegraphics[width=2.5cm]{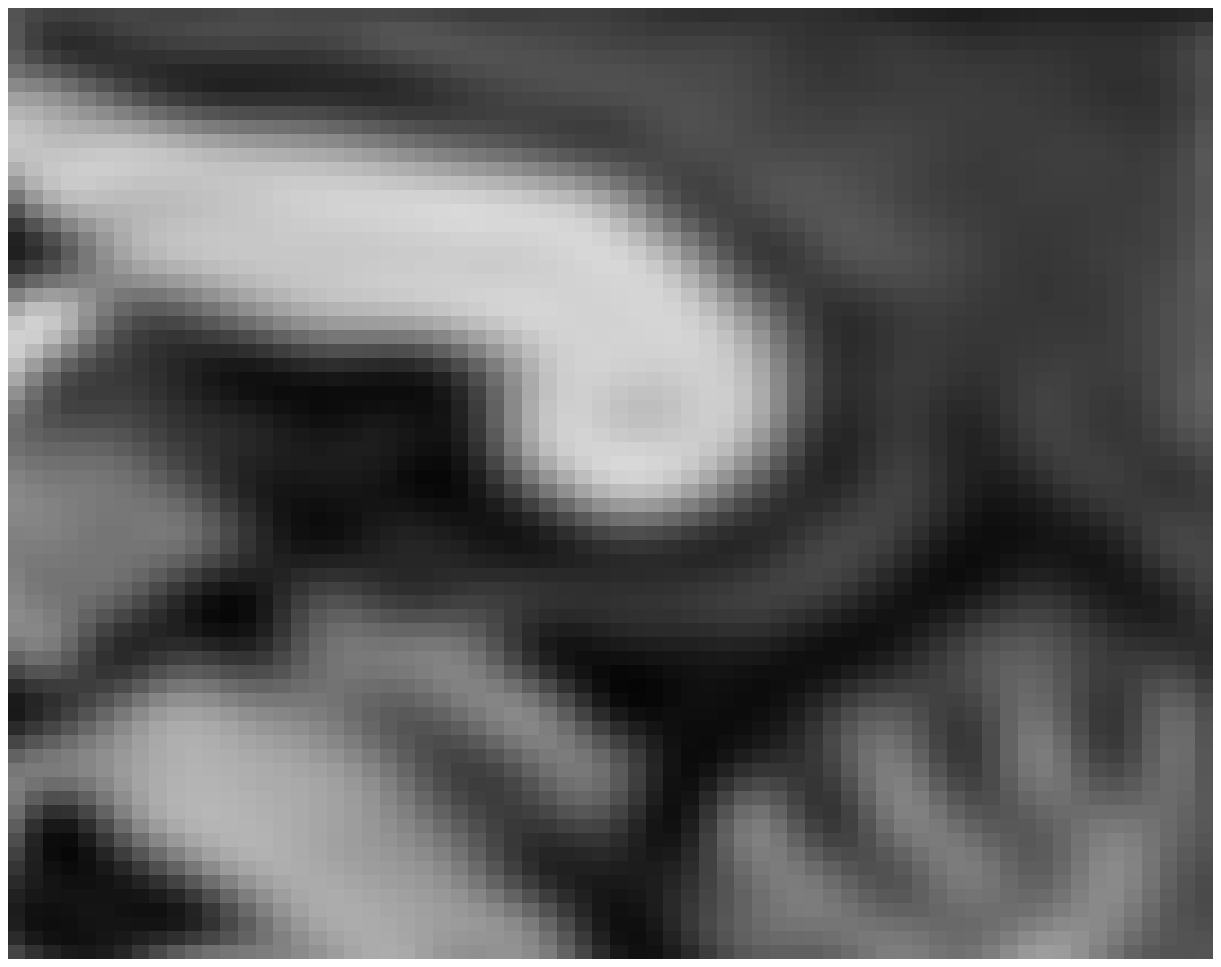} \\
          (b)
        \end{tabular}
        &\hspace{-0.8cm}
        \begin{tabular}[h]{c}
          \includegraphics[width=2.5cm]{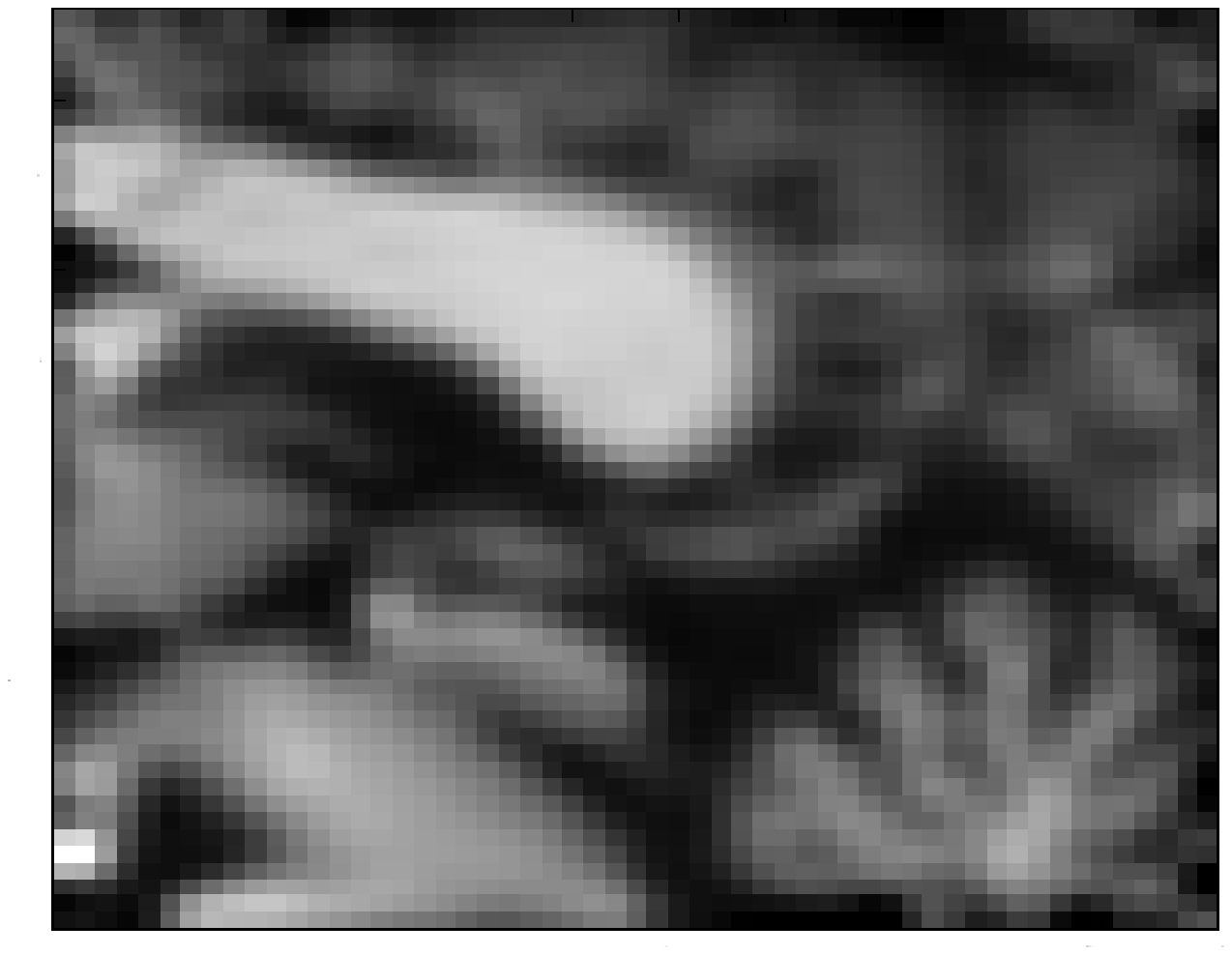}\\
          (c)
        \end{tabular}
        &\hspace{-0.8cm}
        \begin{tabular}[h]{c}
          \includegraphics[width=2.5cm]{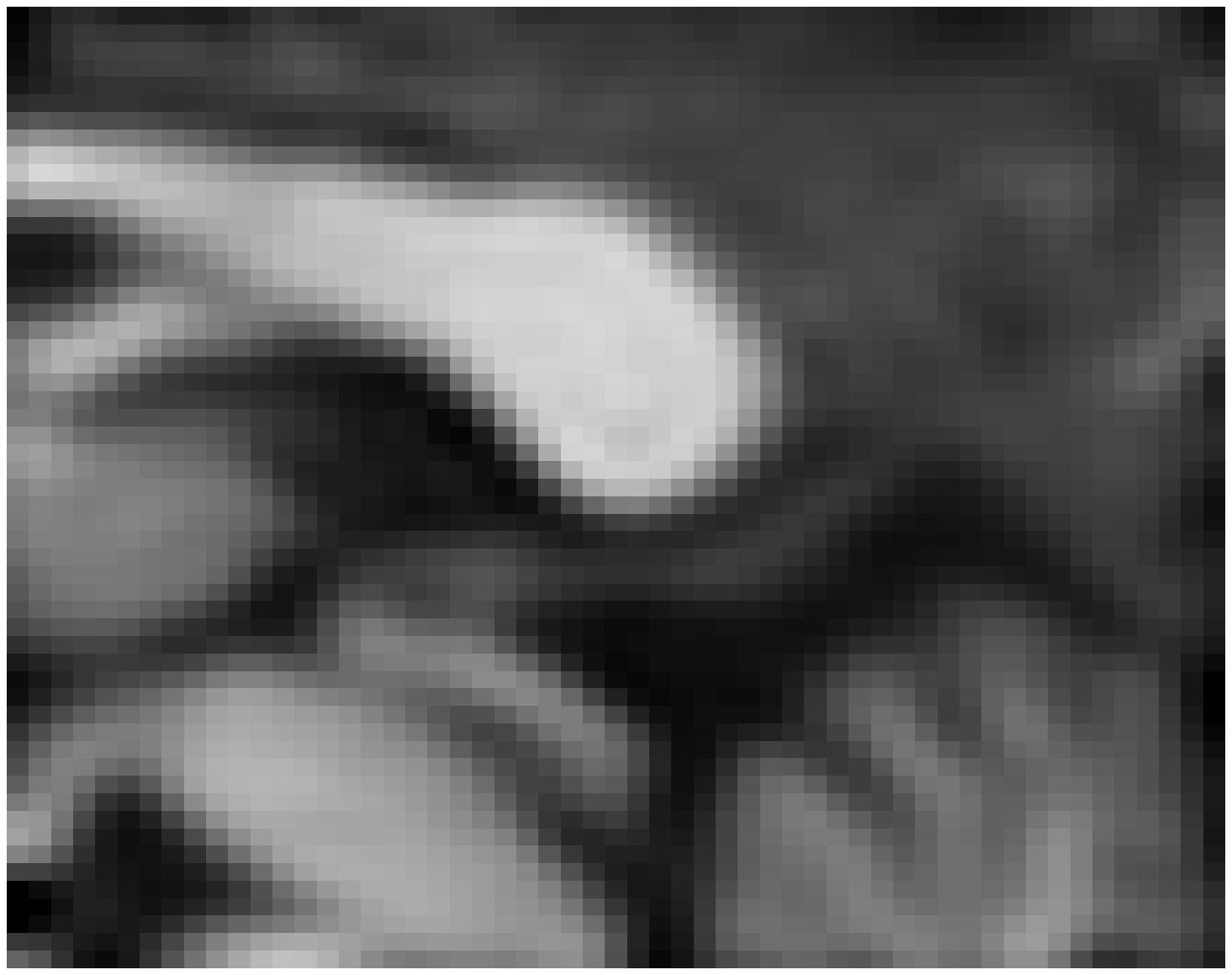}\\
          (d)
        \end{tabular}
        &\hspace{-0.8cm}
        \begin{tabular}[h]{c}
          \includegraphics[width=2.5cm]{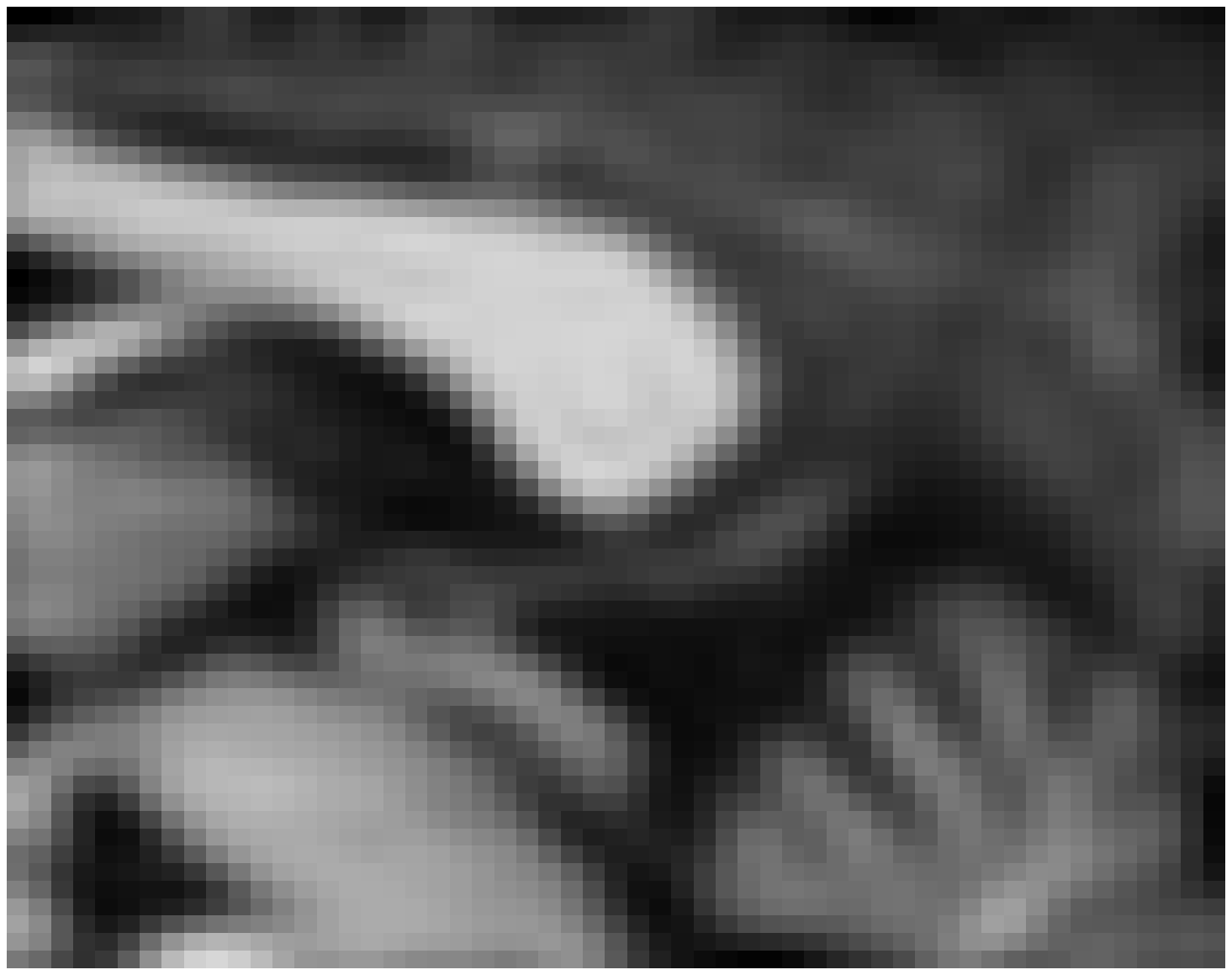}\\
          (e)
        \end{tabular}\\
        &
        &
      \end{tabular}
    \end{tabular}
  \end{minipage}

\caption{First row~: Ten images of the training set representing the splenium and
a part of the cerebellum.
Second row~: Results from the template estimation. (a)~: gray
  level mean image of the 47 images. Templates estimated (b)~: with the FAM (c)~: with the
  stochastic algorithm on the
  simple model (d,e)~: on the two component model. 
}
\label{fig:templatesSP}
\end{figure}
\end{center}

We also test the algorithm on a database which consists in $47$ 2D
medical images. Each of them represents the splenium
(back of the corpus calosum) and a part of the cerebellum. Some of the
training images are shown in Figure \ref{fig:templatesSP} first row.

The results of the estimation are presented in Figure
\ref{fig:templatesSP}, second row.
The first picture presented, (a), is the gray level mean of the $47$
images. The second one, (b), shows the estimated template computed with the
Fast Approximation with Mode Algorithm presented in \cite{AAT} for a
single component model.  This
algorithm is an EM-like algorithm where the E step is simplified. The
posterior distribution of the hidden variable is approximated by a
Dirac distribution on its mode. This yields a deterministic algorithm,
quite simple to implement but with no theoretical convergence
properties.  It  shows a well contrasted
splenium whereas the cerebellum remains a little bit blurry (note that
it is still much better that the simple mean (a)).

This picture has to be compared with picture (c) which gives the
estimated template computed with our algorithm with $\tau_m=1$. The
great improvement from the gray
level mean of the images (a) or the FAM estimation (b) to our
estimations is obvious. In
particular, the splenium is still very contrasted, better localised
and the cerebellum is
reconstructed with several branches. The background presents
 several structures whereas the other estimates are blurry.
 The two anatomical shapes are relevant representants of the
ones observed in the training set.

The estimation has been done while enabling the decomposition of the
database into two components with our SAEM-MCMC algorithm presented
here.
The two estimated templates  are shown
in Figure \ref{fig:templatesSP} (d) and (e). The differences
can be seen in particular on the shape of the splenium where the
boundaries are more or less curved. The thickness of the splenium
varies as well between the two estimates. The position of the
fornix is also different, being closer to the boundary of the
image. The number
of branches in the two cerebella also tends to be different from one
template to the other (4 in the first component and 5 in the second
one).

The estimation suffers from the small number of images we
have. This can be seen in the estimation of the background which is
blurry in both images. To be able to explain the huge variability of
the two anatomical
shapes, more components would be interesting but at the same time more
images required so that the components will not end up empty.

\section{Proof of Theorem \ref{saemlike}}\label{proof}
We recall that in this
section the variances of the components are fixed. This reduces the
parameters $\theta_t$ to $(\alpha_t,\Gamma_{g,t})$ for all $1\leq t\leq
\tau_m$.\\

First let exhibit sufficient statistics for the model.
The complete log-likelihood equals:
\begin{eqnarray*}
 \log q(\bdy,\bdbeta,\bdtau|\eta) 
 &=& \sum\limits_{i=1}^n
\left\{
\log \left[
\left( \frac{1}{2 \pi \si_{\tau_i}^2}\right)^{|\Lambda|/2}
\exp \left(
-\frac{1}{2\si_{\tau_i}^2} \|y_i - K_p^{\beta_i} \alpha_{\tau_i} \|^2
\right)
\right] \right. \\
&&+ \left. \log \left[
\left( \frac{1}{2 \pi}\right)^{k_g} |\Gamma_{g, \tau_i}|^{-1/2}
\exp \left(
-\frac{1}{2} \beta_i^t \Gamma_{g, \tau_i}^{-1} \beta_i
\right)
\right]  + \log (\rho_{\tau_i}) \right\},
\end{eqnarray*}
where $K_p^{\beta} \alpha=z_\beta I_\alpha $ and $\|.\| $ denotes the Euclidean norm. 
This emphasises five sufficient statistics
given in their
matricial form for all $ 1\leq t \leq \tau_m $,
$$
\begin{array}{lll}
  S_{0,t}(\bdbeta,\bdtau)  &=&\sum\limits_{1\leq i\leq n} \mathds{1}_{\tau_i
  = t} \,,\\
S_{1,t}(\bdbeta,\bdtau)  &=&   \sum\limits_{1\leq i\leq n}\mathds{1}_{\tau_i
  = t} \left( K_p^{\beta_i}
\right)^t y_i \,,\\
S_{2,t}(\bdbeta,\bdtau) &=&\sum\limits_{1\leq i\leq n} \mathds{1}_{\tau_i
  = t} \left( K_p^{\beta_i}
\right)^t \left( K_p^{\beta_i}
\right) \,,\\
S_{3,t}(\bdbeta,\bdtau) & = &\sum\limits_{1\leq i\leq n}\mathds{1}_{\tau_i
  = t} \beta_i ^t \beta_i \,,�\\
 \tS_{4,t}(\bdbeta,\bdtau)&=& \sum\limits_{1\leq i\leq n} \mathds{1}_{\tau_i
  = t} \|y_i\|^2 \, .
\end{array}$$
Thus we apply the stochastic approximation at iteration $k$ of the
algorithm leading to:
$$s_{k,m,t}=s_{k-1,m,t}+\Delta_k(S_{m,t}(\bdbeta_k,\bdtau_k)-s_{k-1,m,t})
$$
for $0 \leq m\leq 4$ and rewrite the maximisation step.
The weights and the covariance matrix are updated as follows:
\begin{eqnarray}\label{rhoup}
\rho_{\tau,k} &=& \frac{s_{k,0,\tau} + a_\rho}{n + \tau_m a_\rho} \,,\\
\Gamma_{g,\tau,k} &=& \frac{1}{s_{k,0,\tau}+a_g}
(s_{k,0,\tau} s_{k,3,\tau} + a_g \Gammag) \, .
\end{eqnarray}
The photometric parameters are solution of the
following system:
\begin{equation}  \label{PhotoUpdateClust}
\left\{
\begin{array}{lll}
\alpha_{\tau,k} &=& \left(s_{k,0,\tau} s_{k,2,\tau}
+ \sigma_{\tau,k} ^2  (\Gammap)^{-1}
\right)^{-1}  \left(s_{k,0,\tau} s_{k,1,\tau} +  \sigma_{\tau,k} ^2
(\Gammap)^{-1} \mu_p \right) \,,\\
\\
\sigma^2_{\tau,k} &= &  \frac{1}{s_{k,0,\tau} |\Lambda| +a_p} \left(
s_{k,0,\tau}\left( s_{k,4,\tau} + (\alpha_{\tau,k})^t  s_{k,2,\tau} \alpha_{\tau,k}
- 2  (\alpha_{\tau,k})^t
s_{k,1,\tau}\right) + a_p\sigma_0^2 \right) \, ,
\end{array}\right.
\end{equation}
which can be solved iteratively for each component $\tau$ starting with
the previous values. In this part, all $\sigma^2_{\tau,k} $ are fixed and this
leads to an explicit form for the parameters $ \alpha_{\tau,k} $.\\

We will now apply Theorem \ref{SA} to prove Theorem \ref{saemlike}.

(\textbf{SA0'}) is  satisfied with the set $\Sa$ defined by
$$ \Sa\triangleq \{ \ S \in \mathcal{S} \ |\ 0 \leq S_{0,t}\leq n,\ \| S_{1,t}\| \leq \|\bdy\| ,\
\| S_{2,t}\| \leq n,\ 0\leq S_{3,t},\ 0\leq S_{4,t}\leq \|\bdy\| ^2, \ \forall 1\leq t\leq \tau_m
\   \} \ .$$ Thanks to the convexity of this set, the new value $s_k$
defined as a barycenter
remains in $\mathcal{S}_a$.

Assumption (\textbf{SA1}) is trivially satisfied since
we can choose our step-size sequence $(\Delta_k)_k$.\\

(\textbf{SA2}) holds as already proved in \cite{aktdefmod} for the one
component case
with $w(s) = -l(\hat\e(s))$ such as (\textbf{STAB1'(ii)}) with the
same function $W(s) = -l(\hat\e(s))$. These conditions  imply the
contraction property of the Lyapunov function $w$ and the convergence
of the stochastic approximation under some conditions on the
perturbations. \\

We need to suppose, like in the one component case \cite{aktdefmod},
that the critical
points of our model are in a compact subset of $\mathcal{S}$ which
stands for  (\textbf{STAB1'(i)}). This is an assumption which has to
be considered in a future work.\\

We will now focus on (\textbf{STAB2}) which is the assumption which
gives the control of the perturbations required for the
convergence. \\

We first show the convergence
to zero of the remainder term $|r_k|\mathds{1}_{W(s_{k-1}) \leq M}$ for
any positive integer $M$.
 We denote
by $\pi_k=\pi_{\hat{\e}(s_k)}$ for any $k\geq 0$. We have
$r_k =\mathbb{E}\left[
  \tS   (\bdbeta_k,\la_k) |\mathcal{F}_{k-1} 
\right] - \mathbb{E}_{\lstat_{k-1}} \left[ \tS (\bdbeta,\la)
\right] $ thus,
\begin{eqnarray*}
r_k&=&  \sum\limits_{\la} \int_{\R^N}  \tS(\bdbeta,\la)
\Pi^{\nmck}_{\e_{k-1},\la}(\bdbeta_{0},\bdbeta) \prod_{i=1}^n \int
  p_{\nmck}(\tau_i|\xi_{i},y_i,\e_{k-1}) \prod_{t=1}^{\tau_m}
\prod_{l=1}^{\nmck} \Pi_{\e,t}
(\xi_{t,i}^{(l-1)} ,\xi_{t,i}^{(l)} )d\xi_{t,i}^{(l)}d\bdbeta\\
& &- \sum\limits_{\la} \int_{\R^N}   \tS(\bdbeta,\la)
\lstat_{\e_{k-1}}(\bdbeta,\la)  d\bdbeta\,. \\
\end{eqnarray*}

\noindent We denote by $Q(\xi_{i})d\xi_{i}=\prod\limits_{t=1}^{\tau_m}  \prod\limits_{l=1}^{\nmck} \Pi_{\e,t}
(\xi_{t,i}^{(l-1)} ,\xi_{t,i}^{(l)}
) \ d\xi_{t,i}^{(l)}$ and by\\
 $R_{\nmck}(\la|y,\e_{k-1})= \prod\limits_{i=1}^n \int
  p_{\nmck}(\tau_i|\xi_{i},y_i,\e_{k-1})   \
Q(\xi_{i}) \ d\xi_{i} .$
We can now rewrite
\begin{eqnarray*}
|r_k|&\leq& \left|\sum\limits_{\la} \int_ {\R^N} \tS(\bdbeta,\la) \left[
    \Pi^{\nmck}_{\e_{k-1},\la}(\bdbeta_{0},\bdbeta)R_{\nmck}(\la|\bdy,\e_{k-1}) d\bdbeta-
 \lstat_{\e_{k-1}}(\bdbeta,\la) \right] d\bdbeta\right|\\
&\leq& \sum\limits_{\la}\left|\int_{\R^N} \tS(\bdbeta,\la)  \left[
    \Pi^{\nmck}_{\e_{k-1},\la}(\bdbeta_{0},\bdbeta)-q(\bdbeta|\la,\bdy,\e_{k-1})\right]
 d\bdbeta\right|\left|  R_{\nmck}(\la|\bdy,\e_{k-1})
 \right| \\
&&+\sum\limits_{\la} \left|\int_{\R^N}  \tS(\bdbeta,\la)
  q(\bdbeta|\la,\bdy,\e_{k-1})d\bdbeta\right|
\left|R_{\nmck}(\la|\bdy,\e_{k-1})
 -q(\la|\bdy,\e_{k-1})\right|.
 \end{eqnarray*}
 Denoting $\mathcal{M}_{\e_{k-1}}= \max_{\la} \int_{\R^N}  |\tS(\bdbeta,\la)|
 q(\bdbeta|\la,\bdy,\e_{k-1})d\bdbeta$, we obtain finally
 \begin{eqnarray}
|r_k|\mathds{1}_{W(s_{k-1}) \leq M} &\leq& \sum\limits_{\la}\left|\int_{\R^N}
  \tS(\bdbeta,\la)  \left[
    \Pi^{\nmck}_{\e_{k-1},\la}(\bdbeta_{0},\bdbeta)-q(\bdbeta|\la,\bdy,\e_{k-1})\right]
 d\bdbeta\right| \mathds{1}_{W(s_{k-1}) \leq M}\label{terme1}\\
&&+ \mathcal{M}_{\e_{k-1}}\sum\limits_{\la} \left|R_{\nmck}(\la|\bdy,\e_{k-1})
 -q(\la|\bdy,\e_{k-1})\right|\mathds{1}_{W(s_{k-1}) \leq M}.
 \label{terme2}
\end{eqnarray}
We will first show  that the Gibbs
sampler kernel $\Pi_{\e,\la}$ satisfies a lower bound condition and a
Drift condition
(\textbf{MDRI}) to get its geometric ergodicity (as it has been done in
\cite{aktdefmod}).
\begin{description}
\item[(MDRI)]
For any $s\in \mathcal{S}$ and any $\la\in\T$, $\ntrans_{\hat{\e}(s),\la}$
is irreducible  and
aperiodic. In addition there exists a function $V:\R^{N} \to
[1,\infty[$
such  that  for any $p\geq 1$
and
any compact subset $\Kapa\subset \mathcal{S}$,
 there exist a  set $\texttt{C}$, an integer $m$, constants $0<\kappa<1$, $B>0$
, $\delta>0$ and a probability measure $\nu$ such that

\begin{eqnarray}
\label{6.3}
 \inf\limits_{s\in\Kapa,\la\in\T} \ntrans_{\hat{\e}(s),\la} ^m (\bdbeta,A) &\geq& \delta \nu(A) \ \
 \forall \bdbeta\in \texttt{C}, \forall A \in \mathcal{B}(\R^{N})\,, \\
 \label{6.1}\sup\limits_{s\in\Kapa,\la\in\T} \ntrans_{\hat{\e}(s),\la} ^mV^p(\bdbeta)& \leq &\kappa V^p(\bdbeta) +
 B \mathds{1}_{\texttt{C}}(\bdbeta) \,.
\end{eqnarray}
\end{description}

\begin{notation}
Let
$(e_j)_{1\leq j\leq N}$ be the canonical basis of the $\bdbeta$-space and for any
$1\leq j\leq N$, let
$E_{\e,\la,j} \triangleq  \{\ \bdbeta\in\mathbb{R}^N\ |\ \langle
\bdbeta,e_j\rangle_{\e,\la}=0\}$ be the orthogonal of $\text{Span}\{e_j\}$
and $p_{\e,\la,j}$ be the orthogonal projection on $E_{\e,\la,j}$ i.e.
$$p_{\e,\la,j}(\bdbeta) \triangleq
\bdbeta-\frac{\langle \bdbeta,e_j\rangle_{\e,\la}}{\|e_j\|^2_{\e,\la}}e_j\,,$$
where $\langle \bdbeta,\bdbeta'\rangle_{\e,\la}=\sum_{i=1}^n
\beta_i^t\Gamma_{g,\tau_i}^{-1}\beta_i'$ for 
$\bdbeta$ and  $\bdbeta'$ in $\R^N$ (i.e. the natural dot product
associated with the covariance matrices $(\Gamma_{g,t})_t)$ and $\| .\| _{\eta,
  \tau} $ is the corresponding norm.

  We denote for any $1\leq j\leq N$, $\e\in\Theta� \varrho$ and $\la\in\T$,
 by $\Pi_{\e,\la,j}$ the Markov kernel on $\mathbb{R}^N$ associated with the $j$-th
  Metropolis-Hastings step of the Gibbs sampler on $\mathbb{R}^N$.
We have $\Pi_{\e,\la}=\Pi_{\e,\la,N}\circ\cdots\circ\Pi_{\e,\la,1}$.
\end{notation}

\vspace{0.3cm}

Inequality \eqref{6.3} is equivalent to the existence of a small set
$\texttt{C}$ for the kernel  $\ntrans_{\hat\eta(s),\tau}$ independent of $s\in \Kapa$. We recall here the definition
of a small set:

\vspace{0.3cm}
\begin{Def} (cf. \cite{meyntweedie})
 A set $\mathcal{E} \in \mathcal{B}(\R^N)$ is called a
 \textbf{small set} for the kernel $\ntrans$
 if there exist an integer $m>0$ and a non trivial measure $\nu_m$ on $
 \mathcal{B}(\R^N)$, such that for all $\bdbeta\in \mathcal{E}$,  $B\in
 \mathcal{B}(\R^N)$,
  $ \ntrans^m(\bdbeta,B) \geq \nu_m(B).$

When this holds, we say that $\mathcal{E}$ is $\nu_m$-small.
\end{Def}
\vspace{0.3cm}

We now prove the following lemma:
\vspace{0.3cm}
\begin{lemma}
 Let $\E$ be a compact subset of $\mathbb{R}^N$  and $\Kapa$ be a
 compact subset of $\mathcal{S}$, then $\E$ is a small
set of $\R^N$ for $(\ntrans_{\hat\e(s),\la})_{s\in\Kapa,\la\in\T}$.
\end{lemma}
\begin{proof}
The transition probability kernel of our Markov chain on
$\bdbeta$ is defined as follows~: for coordinate $j$, the kernel is
\begin{multline}\label{eq:kernelj}
\Pi_{\eta,\la,j}(\bdbeta,d\bdz)=\left( \otimes _{m\neq j} \delta_{\bdbeta^m} (d\bdz^m)
\right) � \left[
 \qj(d\bdz^j|\bdbetamj,\eta , \la )r_{j}(\bdbeta^j,d\bdz^j;\bdbetamj,\eta ,\la)
+\right.\\
\left.
\delta_{\bdbeta^j} (d\bdz^j) \int (1- r_{j}(\bdbeta^j,b;\bdbetamj,\eta,\la ) )
 \qj(b|\bdbetamj,\eta,\la) db
\right] \,.
\end{multline}

Then note that there exists $a_c>0$ such that
for any $\e\in \Te�\varrho$, any $\bdbeta\in\mathbb{R}^N$ and any
$b\in\mathbb{R}$, the acceptance rate
  $r_{j}(\bdbeta^j,b; \bdbetamj,\eta,\bdtau)$  is uniformly
lower bounded  by $a_c$ so that for any $1\leq j\leq N$ and any non-negative
  function $f$,
$$\Pi_{\e,\la,j}f(\bdbeta)\geq
a_c \int_\mathbb{R}f(\bdbeta^{-j}+be_j)q_j(b|\bdbeta^{-j},\la,\e)db= a_c \int_\mathbb{R}f(p_{\e,\la,j}(\bdbeta)+ze_j/\|e_j\|_{\e,\la})g_{0,1}(z)dz\,,$$
where $g_{0,1}$ is the probability density function of the standard $\mathcal{N}(0,1)$.
By induction, we have
\begin{equation}
\Pi_{\e,\la}f(\bdbeta)\geq
a_c^N \int_{\mathbb{R}^N}f\left(p_{\e,\la,1,N}(\bdbeta)+\sum_{j=1}^Nz_jp_{\e,\la,j+1,N}(e_j)/\|e_j\|_{\e,\la}\right)\prod_{j=1}^Ng_{0,1}(z_j)dz_j\,,\label{eq:6}
\end{equation}
where $p_{\e,\la,q,r}=p_{\e,\la,r}\circ p_{\e,\la,r-1} \circ  \cdots\circ
  p_{\e,\la,q}$ for any integer $q\leq r$ and $p_{\e,\la,N+1,N}=Id_{\R^N}$.
 Let $A_{\e,\la}\in \mathcal{L}(\mathbb{R}^N)$ be the linear
mapping on $z_1^N=(z_1,\cdots,z_N)$ defined by $$A_{\e,\la}
z_1^N=\sum_{j=1}^Nz_jp_{\e,\la,j+1,N}(e_j)/\|e_j\|_{\e,\la}\,.$$ One easily
checks that for any $1\leq k\leq N$, $\text{Span}\{\
p_{\e,\la,j+1,N}(e_j),\ k\leq j\leq N\}=\text{Span}\{ e_j, k\leq
j\leq N\}$
so that $A_{\e,\la}$ is an invertible mapping. By a change of variable,
we get
$$\int_{\mathbb{R}^N}f\left(p_{\e,\la,1,N}(\bdbeta)+A_{\e,\la}
  z_1^N\right)\prod_{j=1}^Ng_{0,1}(z_j)dz_j=\int_{\mathbb{R}^N}f(u)g_{p_{\e,\la,1,N}(\bdbeta),A_{\e,\la}
  A_{\e,\la}^t}(u)du\,,$$
where $g_{\mu,\Sigma}$ stands for the probability density function of the normal law
  $\mathcal{N}(\mu,\Sigma)$.\\

Since
  $(\e,\la)\to A_{\e,\la}$ is smooth on the set of invertible mappings in
  $(\e,\la)$, we deduce that there exist $c_\Kapa>0$ and $C_\Kapa>0$
  such that $c_\Kapa\text{Id}\leq
  A_{\e,\la} A_{\e,\la}^t\leq \text{Id}/c_\Kapa$ and $g_{p_{\e,\la,1,N}(\bdbeta),A_{\e,\la}
  A_{\e,\la}^t}(u)\geq C_\Kapa g_{p_{\e,\la,1,N}(\bdbeta),\text{Id}/c}(u)$ uniformly
  for $\e=\hat\e(s)$ with $s\in\mathcal{K}$ and $\la\in\T$. Assuming that
  $\bdbeta\in\E$, since $\e\to p_{\e,\la,1,N}$ is smooth and $\E$ is
  compact, we have $\sup_{\bdbeta\in\E,\e=\hat\e(s),\ s\in
  \mathcal{K},\la\in\T}\|p_{\e,\la,1,N}(\bdbeta)\|<\infty$ so that there exist
other constants $C_\Kapa>0$
  and $c_\Kapa>0$ such that for any
  $(u,\bdbeta)\in\mathbb{R}^N� \E$ and any $\e=\hat\e(s),\
  s\in\mathcal{K},\la\in\T$
\begin{equation}
g_{p_{\e,\la,1,N}(\bdbeta),A_{\e,\la}
  A_{\e,\la}^t}(u)\geq C_\Kapa g_{0,\text{Id}/c_\Kapa}(u)  .\label{eq:5}
\end{equation}
  Using \eqref{eq:6} and \eqref{eq:5}, we deduce that for any $A$
$\Pi_{\e,\la}(\bdbeta,A)\geq C_\Kapa a_c^N \nu(A)\,,$
with $\nu$ equal to the density of the normal law
$\mathcal{N}(0,\text{Id}/c_\Kapa)$.
  This yields the
existence of the small set as well as equation
(\ref{6.3}).
\end{proof}
\vspace{0.3cm}

This property also implies the $\phi$-irreducibility of the Markov
Chain generated by $\ntrans_{\e,\la}$.
Moreover, the existence of a $\nu_1$-small set implies the
aperiodicity of the chain (cf. \cite{meyntweedie}).
\vspace{0.3cm}

Now consider the Drift condition (\ref{6.1}).\\

We set $V:\mathbb{R}^N\to [1,+\infty [$ as the following function
$V(\bdbeta) = 1 + \|\bdbeta\|^2,
$
where $\| \cdot \|$ denotes the Euclidean norm.
Define for
any $g:\R^N \to
\R^{n_s}$ the norm
$\|g\|_V = \sup\limits_{ \bdbeta \in \R^N } \frac{\|g(\bdbeta)\|}{V(\bdbeta)}$
and the functional space $\mathcal{L}_V=\{g:\R^N \to
\R^{n_s} \  | \ \|g\|_V < +\infty\}$.
For any $\e\in\Te� \varrho$ and any $\la\in\T$, we introduce a $(\e,\la)$
dependent function $V_{\e,\la}(\bdbeta)\triangleq 1+\| \bdbeta\|^2
_{\e,\la}$.
\vspace{0.5cm}

\begin{lemma}
\label{lem:a1}
 Let $K$ be a compact subset of $\Te�\varrho$. For any integer $p\geq 1$, there
 exist $0< \rho<1$ and
  $C>0$ such that for any $\e\in K$, any $\la\in\T$,    any
  $\bdbeta\in\mathbb{R}^N$ we
  have
  $$\Pi_{\e,\la} V_{\e,\la}^p(\bdbeta)\leq \rho V_{\e,\la}^p(\bdbeta)+C\,.$$
\end{lemma}

\begin{proof}
The proposal distribution for $\Pi_{\e,\la,j}$ is given by $\qj(\bdbeta^j \ |\
\bdbetamj,\la,\bdy,\e)\stackrel{\text{law}}{=}p_{\e,\la,j}(\bdbeta)+
\frac{z}{\|e_j\|_{\eta,\tau}}
e_j$ , where $z\sim \mathcal{N}(0,1)$.
Then,  for any $\bdbeta\in\mathbb{R}^N$ and  any
measurable set $A\in\mathcal{B}(\mathbb{R}^N)$, there exists $a_{\eta,
  \bdtau,j}(\bdbeta)$
uniformly
bounded from below
by  $a_c>0$ such that
$$\Pi_{\e,\la,j}(\bdbeta,A)=(1-a_{\eta,
  \bdtau,j}(\bdbeta) )\mathds{1}_A(\bdbeta)+a_{\eta,
  \bdtau,j}(\bdbeta)
\int_\mathbb{R}\mathds{1}_A\left(p_{\e,\la,j}(\bdbeta)+\frac{z}{\|e_j\|_{\eta,\tau}}e_j\right)g_{0,1}
(dz)\,,$$

Since $\langle
p_{\e,\la,j}(\bdbeta),e_j\rangle_{\e,\la}=0$, we get
$V_{\e,\la}\left(p_{\e,\la,j}(\bdbeta)+\frac{z}{\|e_j\|_{\eta,\tau}}e_j\right)=
V_{\e,\la}(p_{\e,\la,j}(\bdbeta))+z^2$. We deduce that there exists $C$ such
that for any $\bdbeta\in\mathbb{R}^N$~:
\begin{eqnarray*}
  \Pi_{\e,\la,j} V_{\e,\la} ^p(\bdbeta)&=&
  (1-a_{\eta,
  \bdtau,j}(\bdbeta))V^p_{\e,\la}(\bdbeta)+a_{\eta,
  \bdtau,j}(\bdbeta)
\int_\mathbb{R} \left(V_{\e,\la}
(p_{\e,\la,j} (\bdbeta))+z^2\right)^pg_{0,1}(dz)\label{eq:1}\\
&\leq&
  (1-a_{\eta,
  \bdtau,j}(\bdbeta))V^p_{\e,\la}(\bdbeta)
+a_{\eta,
  \bdtau,j}(\bdbeta) � \\
&&\left(V^p_{\e,\la}(p_{\e,\la,j}(\bdbeta))+V^{p-1}_{\e,\la}(p_{\e,\la,j}(\bdbeta))
\int_\mathbb{R}(1+z^2)^{p}g_{0,1}(dz)\right)\\
&\leq&
  (1-a_{\eta,
  \bdtau,j}(\bdbeta))V^p_{\e,\la}(\bdbeta)+a_{\eta,
  \bdtau,j}(\bdbeta)V^p_{\e,\la}(p_{\e,\la,j}
(\bdbeta))+C V^{p-1}_{\e,\la}(p_{\e,\la,j}(\bdbeta)) \,.
\end{eqnarray*}
We have used in the last inequality the fact that a Gaussian variable
has bounded
moment of any order. Since $a_{\e,\la,j}(\bdbeta)\geq a_c$ and
$\|p_{\e,\la,j}(\bdbeta)\|_{\e,\la}\leq \|\bdbeta\|_{\e,\la}$ ($p_{\e,\la,j}$ is an
orthonormal projection for the dot product $\langle \cdot,\cdot
\rangle_{\e,\la}$), we get that for any $\varepsilon>0$, there exists
$C_{K,\varepsilon}$ such that for any $\bdbeta\in\mathbb{R}^N$ and $\e\in K,\la\in\T$
$$  \Pi_{\e,\la,j} V^p_{\e,\la}(\bdbeta)\leq
(1-a_c)V^p_{\e,\la}(\bdbeta)+(a_c+\varepsilon
)V_{\e,\la}^p(p_{\e,\la,j}(\bdbeta))+C_{K,\varepsilon }\,.$$
By induction,  we get
$$\Pi_{\e,\la} V^p_{\e,\la}(\bdbeta)\leq
\sum_{u\in\{0,1\}^N}\prod_{j=1}^N(1-a_c)^{1-u_j}(a_c+\varepsilon)^{u_j}
V^p_{\e,\la}(p_{\e,\la,u}(\bdbeta))+\frac{C_{K,\varepsilon } }{\varepsilon } \left( (1+\varepsilon)^{N+1} -1\right)\,,$$
where $p_{\e,\la,u}=((1-u_N)\text{Id}+u_Np_{\e,\la,N})\circ\cdots\circ
((1-u_1)\text{Id}+u_1p_{\e,\la,1})$.

Let
$p_{\e,\la}=p_{\e,\la,N}\circ\cdots\circ
p_{\e,\la,1}$ and note that $p_{\e,\la,u}$ is  contracting so that

$$\Pi_{\e,\la} V^p_{\e,\la}(\bdbeta)\leq b_{c,\varepsilon
}V^p_{\e,\la}(\bdbeta)+(a_c+\varepsilon )^N
V^p_{\e,\la}(p_{\e,\la}(\bdbeta))+\frac{C_{K,\varepsilon }  }{\varepsilon  } \left( (1+\varepsilon )^{N+1}\right)$$
for $b_{c,\varepsilon }=\left(\sum_{u\in\{0,1\}^N,\ u\neq
  \mathbf{1}}\prod_{j=1}^N(1-a_c)^{1-u_j}(a_c+\varepsilon)^{u_j}\right)$. To
end the proof, we need to check that $p_{\e,\la}$ is strictly
contracting uniformly on $K$. Indeed,
$\|p_{\e,\la}(\bdbeta)\|_{\e,\la}=\|\bdbeta\|_{\e,\la}$ implies
that $p_{\e,\la,j}(\bdbeta)=\bdbeta$ for any $1\leq j\leq N$ so that $\langle
\bdbeta,e_j\rangle_{\e,\la}=0$ and $\bdbeta=0$ since $(e_j)_{1\leq j\leq N}$ is a
basis. Using the continuity of the norm of $p_{\e,\la}$ and
the compactness of $K$, we deduce that there exists $0<\rho_K<1$ such
that $\|p_{\e,\la}(\bdbeta)\|_{\e,\la}\leq \rho_K\|\bdbeta\|_{\e,\la}$ for any
$\bdbeta \in \R^N$,
$\e\in K$ and any $\la\in\T$. Changing $\rho_K$ for $1>\rho'_K>\rho_K$ we get
$(1+\rho_K^2\|\bdbeta\|_{\e,\la}^2)^p\leq {\rho'}_K^{2p}(1+\|\bdbeta\|_{\e,\la}^2)^p+C_K$ for
some uniform constant $C_K$ so that
$$\Pi_{\e,\la}V^p_{\e,\la}(\bdbeta)\leq b_{c,\varepsilon
}V^p_{\e,\la}(\bdbeta)+{\rho'}_K^{2p}(a_c+\varepsilon )^N
V^p_{\e,\la}(\bdbeta)+C_{K,\varepsilon}.$$
Since we have  $ \inf\limits_{\varepsilon >0} \left\{b_{c,\varepsilon }+{\rho'}_K^{2p}(a_c+\varepsilon
  )^N\right\}<1$  the result
is straightforward.
\end{proof}
\vspace{0.3cm}
\begin{lemma}
\label{lem:2}
  For any compact set $K\subset \Te�\varrho$, any integer $p\geq 0$, there exist
$0<\rho< 1$, $C>0$ and a positive integer $m_0$ such that $\forall m \geq m_0$ , $\forall
\e \in K $, $\forall \bdbeta \in \T $
$$\Pi_{\e,\la}^mV^p(\bdbeta)\leq \rho V^p(\bdbeta)+C\,.$$
\end{lemma}
\vspace{0.3cm}
\begin{proof}
  Indeed, there exist $0\leq c_1\leq c_2$ such that $c_1V(\bdbeta)\leq
  V_{\e,\la}(\bdbeta)\leq c_2 V(\bdbeta)$ for any $(\bdbeta,\e,\la)\in\mathbb{R}^N�
  K�\T$. Then, using the previous lemma, we have $\ntrans^m_{\e,\la} V^p(\bdbeta)\leq
  c_1^{-p}\ntrans^m_{\e,\la} V_{\e,\la}^p(\bdbeta)\leq c_1^{-p}(\rho^m
  V^p_{\e,\la}(\bdbeta)+C/(1-\rho))\leq (c_2/c_1)^p(\rho^m
  V^p(\bdbeta)+C/(1-\rho))$. Choosing $m$ large enough for $
  (c_2/c_1)^p\rho^m<1$ gives the result.
\end{proof}
\vspace{0.3cm}

This finishes the proof of (\ref{6.1}) and in the same time the
(\textbf{MDRI}). \\

 Thanks to this property we can use the following
proposition (cf. \cite{meyntweedie}, \cite{andrieumoulinespriouret} Proposition
 B1) and lemma applied to every sequence $(\xi_{t,i}^{(l)})_l$ with stationary
distribution $q(\cdot|y_i,t,\eta)$ for all $1\leq t \leq \tau_m$ and all $1\leq i \leq n$.
\vspace{0.3cm}
\begin{prop}
\label{prop:ergodic}
Suppose that $\ntrans$ is irreducible and aperiodic and that
$\ntrans^m(\bdbeta_0,.) \geq \mathds{1}_\smallset(\bdbeta_0) \delta\nu(.)$
for a set $\smallset\in \B(\R^N)$, some integer $m$ and $\delta>0$ and that
there is a Drift condition to $\smallset$ in the sense that, for some
$0<\kappa<1$,
$B>0$ and a function $V: \R^N \to [1,+\infty[,$
\begin{equation*}
 \ntrans V (\bdbeta_0) \leq \kappa V(\bdbeta_0 ) \ \forall \bdbeta_0 \nin
 \smallset \ and \
 \sup\limits_{\bdbeta_0\in \smallset} (V(\bdbeta_0 ) + \ntrans V (\bdbeta_0)) \leq B\,.
 \end{equation*}
Then, there exist constants $K$ and $0<\rho<1$, depending only upon
$m,\delta,\kappa,B$, such that,
for all $\bdbeta_0 \in \R^N$, and all $g \in \mathcal{L}_V$
\begin{equation*}
\| \ntrans^ng(\bdbeta_0) -\lstat(g) \|_V \leq K \rho^n \|g\|_V \, .
\end{equation*}
 \end{prop}
\begin{lemma}
\label{lemmeB2}
Assume that there exist an integer $m$ and constants $0<\kappa<1$ and
$\varsigma>0$ and a set $\smallset$ such that
\begin{equation*}
 \ntrans^m V (\bdbeta_0) \leq \kappa V(\bdbeta_0 ) \ \forall \bdbeta_0 \nin
 \smallset \ and \
\ntrans V (\bdbeta_0) \leq \varsigma V(\bdbeta_0 ) \ \forall \bdbeta_0 \in \R^N \
\end{equation*}
for some function $V: \R^N \to [1,+\infty[$. Then there exists a function
$\tilde V$ and constants $0<\rho<1, c>0$ and $C>0$, depending only upon
$m,\kappa,\varsigma$, such that,
\begin{equation*}
 \ntrans \tilde V (\bdbeta_0) \leq \rho \tilde V(\bdbeta_0 ) \ \forall \bdbeta_0 \nin
 \smallset \ and \
cV \leq \tilde V  \leq CV \,.
\end{equation*}
\end{lemma}

\begin{proof} 
  Define
  \begin{equation*}
    \tilde{V} = \sum\limits_{j=1}^ m \kappa ^{1-j/m} \ntrans^{j-1} V \,.
  \end{equation*}
For $\bdbeta_0 \notin \texttt{C} $, we have
\begin{eqnarray*}
  \ntrans \tilde V (\bdbeta_0) &\leq &\sum\limits_{j=1}^ {m-1} \kappa ^{1-j/m}
  \ntrans^{j} V(\bdbeta_0)  + \kappa V(\bdbeta_0)\\
& \leq  & \kappa^{1/m} \tilde V (\bdbeta_0)\,.
\end{eqnarray*}
Therefore we obtain~:
\begin{equation*}
 \kappa ^{1-1/m} V \leq  \tilde V \leq \left(\sum\limits_{j=1}^ m \kappa ^{1-j/m} \zeta^{j-1} \right) V\,.
\end{equation*}
This ends the proof of Lemma \ref{lemmeB2}.
\end{proof}

\vspace{0.3cm}
Thus, applying the Proposition \ref{prop:ergodic} and Lemma
\ref{lemmeB2} to the Drift conditions of  Lemmas \ref{lem:a1} and
\ref{lem:2}, we get that  each Gibbs
sampler kernel $\Pi_{\e,\la}$ is geometrically  ergodic. \\

Let us now go back to the convergence of the first part of the
residual term \eqref{terme1} towards $0$.\\

We use the term $\mathds{1}_{W(s_{k-1}) \leq M}$ to show that
the parameters $\e_{k-1}$ are constrained to move in a compact set of $\Theta�\varrho$.
We show first that the observed log-likelihood $l$ tends to minus
infinity as the parameters tend to the boundary of
$\Theta�\varrho$. Equation (\ref{eqmodel}) implies that for any
$\te\in\Te$ we have:
\begin{equation*}
q(y_i|\beta_i,\tau_i,\alpha,\sigma^2)q(\beta_i|\Gamma_{g, \tau_i})\leq (2\pi\sigma^2)^{-|\Lambda|/2}(2\pi)^{-k_g}|\Gamma_{g, \tau_i}|^{-1/2}\exp \left(
-\frac{1}{2} \beta_i^t \Gamma_{g, \tau_i}^{-1} \beta_i
\right),
\end{equation*}
so that denoting $C$ as a constant:
\begin{multline*}
\log(q(\bdy,\e))\leq \sum_{i=1}^n \left[ -\frac{a_g}{2}  \langle
\Gamma_{g,\tau_i}^{-1},\Sigma_g\rangle_F+ \frac{1+a_g}{2}\log
|\Gamma_{g,\tau_i}^{-1}|
-\frac{a_p\sigma_0^2}{2\sigma_{\tau_i}^2}-\frac{|\Lambda|+a_p}{2}\log(\sigma_{\tau_i}^2)
\right.\\
\left. -\frac{1}{2}( \alpha_{\tau_i}-\mu_p)^t\Sigma_p^{-1}(
  \alpha_{\tau_i}-\mu_p)-a_\rho \log \rho_{\tau_i}\right]+C .
\end{multline*}
It was shown in \cite{AAT} that we have
$\lim\limits_{||\Gamma||+||\Gamma^{-1}||\to \infty}-\frac{a_g}{2}  \langle
\Gamma^{-1},\Sigma_g\rangle_F+ \frac{1+a_g}{2}\log
|\Gamma^{-1}|=-\infty$ \\
and
$\lim\limits_{\|\alpha\|\to
 \infty}-\frac{1}{2}( \alpha-\mu_p)^t\Sigma_p^{-1}(
  \alpha-\mu_p)=-\infty$.
Moreover, we have
 $ \lim_{\rho\to 0} \log (\rho)=-\infty\,,
$
so we get
 $ \lim_{\e\to \partial(\Te � \varrho)} \log q(\bdy,\e)=-\infty\,,
$
 which ensures that for all $M>0$ there exists $\m>0$ such
that $\|\alpha_t\|\geq \m$ or
$||\Gamma_t||+||\Gamma_t^{-1}||\geq \m$  or $\rho_t\leq
\frac{1}{\m}$ implies $-l(\e)\geq M$.

So $W(s_{k-1})\leq M$
implies that for all $1  \leq t  \leq \tau_m$ we have $\|\alpha_t\|\leq \m$,
$||\Gamma_t||+||\Gamma_t^{-1}||\leq \m$ and $\frac{1}{\m}\leq \rho_t\leq 1-
\frac{1}{\m}$ because $\sum_{t=1}^{\tau_m} \rho_t=1$.

Let us denote by
$\mathcal{V}_\m=\Te_\m^{\tau_m}� \left\{(\rho_t)_{1\leq t\leq \tau_m} \in \left[\frac{1}{\m},1-\frac{1}{\m}\right]^{\tau_m} \ | \
\sum_{t=1}^{\tau_m} \rho_t=1\right\} \,,
$
where
\begin{equation*}
\Te_\m=
\left\{\theta=(\alpha,\Gamma_g)\ |\
\alpha\in\mathbb{R}^{k_p}    \,,
\ \Gamma_g \in \Syme\   | \ \|\alpha\| \leq \m,\frac{1}{\m}\leq
||\Gamma_g|| \leq \m\right\}.
\end{equation*}
 So there exists a compact set
$\mathcal{V}_\m$ of $\Theta�\varrho$ such that $W(s_{k-1})\leq M$ implies
$\hat\e(s_{k-1})\in~\mathcal{V}_\m$
and the first term (\ref{terme1}) can be bounded as follows:
\begin{eqnarray*}
\sum\limits_{\la}\left|\int_{\R^N} \tS(\bdbeta,\la)  \left[
    \Pi^{\nmck}_{\e_{k-1},\la}(\bdbeta_{0},\bdbeta)-q(\bdbeta|\la,\bdy,\e_{k-1})\right]
 d\bdbeta\right|\mathds{1}_{W(s_{k-1}) \leq M}\\
\leq \sum\limits_{\la} \sup_{\e\in\mathcal{V}_\m}\left|\int_{\R^N} \tS(\bdbeta,\la)  \left[
    \Pi^{\nmck}_{\e,\la}(\bdbeta_{0},\bdbeta)-q(\bdbeta|\la,\bdy,\e)\right]
 d\bdbeta\right|.
 \end{eqnarray*}
 Since for each $\la$ the function $\bdbeta \to \tS(\bdbeta,\la) $ belongs to $\mathcal{L}_V$, since we have proved that each transition kernel $\Pi_{\e,\la}$
 is geometrically ergodic and since the set $\mathcal{V}_\m$ is
 compact, we can deduce that  the first term (\ref{terme1})
 converges to zero as $\nmck$ tends to infinity.\\

We now consider
 the second term (\ref{terme2}).
We first need  to prove that $\mathcal{M}_{\e_k}\mathds{1}_{W(s_{k-1}) \leq M}$ is uniformly bounded that is to say the integral of the sufficient statistics
are  uniformly bounded on $\{W(s_{k-1}) \leq M\}$ ; we only need
to focus on the sufficient statistic which is not bounded itself: let
$(j,m) \in \{ 1,...,2k_g\}^2$:
\begin{eqnarray*}
  \int
 | \bdbeta^j\bdbeta^m|
  q(\bdbeta|\la,\bdy,\e_{k-1})
    d\bdbeta \mathds{1}_{\e_{k-1} \in \mathcal{V}_\m}&\leq& \int  |\bdbeta^j\bdbeta^m |
    \frac{q(\bdbeta,\bdtau,\bdy,\e_{k-1})}{q(\bdtau,\bdy,\e_{k-1})}d\bdbeta
    \mathds{1}_{\e_{k-1} \in \mathcal{V}_\m}\\
&\leq & \frac{C(\mathcal{V}_\m)}{q(\bdtau,\bdy,\e_{k-1})} \int | \bdbeta^j\bdbeta^m |
 \exp\left(-\frac{1}{2}
\bdbeta^t\hat\Gamma_{g,\la,{k-1}}^{-1} \bdbeta\right) d\bdbeta\\
&\leq&
C(\mathcal{V}_\m) \int  Q(\bdbeta,\hat\Gamma_{g,\la,{k-1}})  \exp \left( -\frac{1}{2}
\| \bdbeta\|^2
  \right)d\bdbeta
 < \infty\,,
\end{eqnarray*}
where $C(\mathcal{V}_\m)$ is a constant depending only on the set $\mathcal{V}_\m$, $\hat\Gamma_{g,\bdtau}$ is the diagonal block matrix with all the
$\Gamma_{g,\tau_i}$ given by the label vector $\bdtau$ and  we have changed
the variable in the last inequality and $Q$ is a
quadratic form in $\bdbeta$ whose coefficients are continuous functions
of elements of the matrix $\Gamma_g$. So we obtain that for all $M>0$ there exists $\m>0$ such that for all integer $k$ we have:
$\mathcal{M}_{\e_k}\mathds{1}_{W(s_{k-1}) \leq M} \leq C(\mathcal{V}_\m).$

We now prove the convergence to $0$ of the second term of the product
involved in (\ref{terme2}).
Let us denote by $\mathcal{R}_{\la,\bdy,k}$ for the term $\left|R_{\nmck}(\la|\bdy,\e_{k-1})
 -q(\la|\bdy,\e_{k-1})\right|$. Thus we have:
\begin{eqnarray*}
\mathcal{R}_{\la,\bdy,k}&=&
\left|\prod_{i=1}^n \int p_{\nmck}(\tau_i|\xi_{i},y_i,\e_{k-1})Q(\xi_i)d\xi_i
    -\prod_{i=1}^n q(\tau_i|y_i,\e_{k-1})\right|\\
&\leq &  \sum_{i=1}^n \left|\int
  p_{\nmck}(\tau_i|\xi_{i},y_i,\e_{k-1})  Q(\xi_i)d\xi_i   -q(\tau_i|y_i,\e_{k-1})\right|\\
&\leq  & \sum_{i=1}^n \int \left|p_{\nmck}(\tau_i|\xi_{i},y_i,\e_{k-1})-q(\tau_i|y_i,\e_{k-1})\right|Q(\xi_i)d\xi_i\\
&\leq &\sum_{i=1}^n \int \left|\frac{S_{\nmck}(\tau_i,y_i|\xi_{\tau_i,i},\e_{k-1})}{\sum_s S_{\nmck}(s,y_i|\xi_{s,i},\e_{k-1}) }
-\frac{q(\tau_i,y_i|\e_{k-1})}{q(y_i|\e_{k-1})}\right|Q(\xi_i)d\xi_i \,,
\end{eqnarray*}
where we denote by $S_\nmc(t,y_i|\xi_{t,i},\e)$ the quantity $\left( \frac{1}{\nmc} \sum\limits_{l=1}^\nmc
 \left[
   \frac{\densMC(\xi^{(l)} _{t,i})}{q(y_i,\xi^{(l)} _{t,i},t|\e)}
 \right]
 \right)^{-1}$.

We write each term of this sum as follows:
\begin{multline*}
\frac{S_{\nmck}(\tau_i,y_i|\xi_{\tau_i,i},\e_{k-1})}{\sum\limits_{s=1}^{\tau_m}
  S_{\nmck}(s,y_i|\xi_{s,i},\e_{k-1}) }
-\frac{q(\tau_i,y_i|\e_{k-1})}{q(y_i|\e_{k-1})}= \\
\frac{S_{\nmck}(\tau_i,y_i|\xi_{\tau_i,i},\e_{k-1})( q(y_i|\e_{k-1})-
\sum\limits_{s=1}^{\tau_m}
S_{\nmck}(s,y_i|\xi_{s,i},\e_{k-1}))}{q(y_i|\e_{k-1})\sum\limits_{s=1}^{\tau_m}
S_{\nmck}(s,y_i|\xi_{s,i},\e_{k-1}) }
 \\
+ \frac{(S_{\nmck}(\tau_i,y_i|\xi_{\tau_i,i},\e_{k-1})-
q(\tau_i,y_i|\e_{k-1}))\sum\limits_{s=1}^{\tau_m}
S_{\nmck}(s,y_i|\xi_{s,i},\e_{k-1})
}{q(y_i|\e_{k-1})\sum\limits_{s=1}^{\tau_m}
  S_{\nmck}(s,y_i|\xi_{s,i},\e_{k-1}) } \,.
 \end{multline*}
Denoting by $\mathcal{T}_i$ the set of $\tau_m+1$ integers $\{1,\cdots,\tau_m\}\cup
\{\tau_i\}$, we obtain finally:
\begin{eqnarray*}
\mathcal{R}_{\la,\bdy,k}
&\leq& \sum_{i=1}^n \frac{1}{q(y_i|\e_{k-1})}\sum_{s \in \mathcal{T}_i} \int \left|S_{\nmck}(s,y_i|\xi_{s,i},\e_{k-1}) -q(s,y_i|\e_{k-1})\right|Q(\xi_i)d\xi_i \,.
\end{eqnarray*}
Defining  the event
 $A_{k,i,t}=\{\left| S_{\nmck}(t,y_i|\xi_{t,i},\e_{k-1}) -q(t,y_i|\e_{k-1})   \right|>\zeta_{k}\} $
for some positive sequence $(\zeta_k)_k$, we get:
\begin{eqnarray*}
\mathcal{R}_{\la,\bdy,k}&\leq& \sum_{i=1}^n \frac{1}{q(y_i|\e_{k-1})}\sum_{s \in
  \mathcal{T}_i}  \int_{ A_{k,i,s}}
\left|S_{\nmck}(s,y_i|\xi_{s,i},\e_{k-1})
  -q(s,y_i|\e_{k-1})\right|Q(\xi_i)d\xi_i \\
&&+\sum_{i=1}^n \frac{1}{q(y_i|\e_{k-1})}\sum_{s \in \mathcal{T}_i} \int_{ A_{k,i,s}^c}
\left|S_{\nmck}(s,y_i|\xi_{s,i},\e_{k-1})
  -q(s,y_i|\e_{k-1})\right|Q(\xi_i)d\xi_i \,.
\end{eqnarray*}
So we deduced that:
\begin{eqnarray*}
\mathcal{R}_{\la,\bdy,k}&\leq& \sum_{i=1}^n \frac{1}{q(y_i|\e_{k-1})}\sum_{s \in
  \mathcal{T}_i}  (\sup_\xi S_{\nmck}(s,y_i|\xi_{s,i},\e_{k-1})
+q(s,y_i|\e_{k-1}))P( A_{k,i,s}) \\
&&+\left(\sum_{i=1}^n \frac{1}{q(y_i|\e_{k-1})}\right) (\tau_m+1)\zeta_k \\
&\leq& \sum_{i=1}^n \left(\frac{\sup_{\xi,s} S_{\nmck}(s,y_i|\xi_{s,i},\e_{k-1}) }{q(y_i|\e_{k-1})}+1\right)\left(\sum_{s \in \mathcal{T}_i}  P( A_{k,i,s})+ P( A_{k,i,\tau_i})\right)\\
&&+\left(\sum_{i=1}^n \frac{1}{q(y_i|\e_{k-1})}\right) (\tau_m+1)\zeta_k \,.
\end{eqnarray*}

 Assuming $\zeta_k< \min_{i,t} q(t,y_i|\e_{k-1})$, we obtain:
\begin{eqnarray*}
 P(A_{k,i,t}^c)&=&P(\left|S_{\nmck}(t,y_i|\xi_{t,i},\e_{k-1}) -q(t,y_i|\e_{k-1})\right|\leq\zeta_k)\\
  &\geq &P\left(\left|      \frac{1}{S_{\nmck}(t,y_i|\xi_{t,i},\e_{k-1}) }
-\frac{1}{q(t,y_i|\e_{k-1})} \right |\leq
\frac{\zeta_k}{q(t,y_i|\e_{k-1})(q(t,y_i|\e_{k-1})+\zeta_k)}\right)\\
 &\geq &P\left(\left|      \frac{1}{S_{\nmck}(t,y_i|\xi_{t,i},\e_{k-1}) }
-\frac{1}{q(t,y_i|\e_{k-1})} \right |\leq
\frac{\zeta_k}{2q(t,y_i|\e_{k-1})^2}\right)\,.
\end{eqnarray*}
Using the first inequality of Theorem 2 of \cite{doreazhao}, we get:
$P(A_{k,i,t}) \leq  c_1 \exp \left(-c_2 \frac{\nmck \zeta_k^2}{q(t,y_i|\e_{k-1})^4}\right),$
where $c_1$ and $c_2$ are independent of $k$ since $(\eta_k)$ only
moves in a compact set $\mathcal{V}_\m $  thanks to the condition $\mathds{1}_{W(s_{k-1}\leq
M)}$.
This yields:
\begin{multline*}
\mathcal{R}_{\la,\bdy,k}
 \leq c_1 \sum_{i=1}^n \left(\frac{\sup_{\xi,s}
     S_{\nmck}(s,y_i|\xi_{s,i},\e_{k-1})
   }{q(y_i|\e_{k-1})}+1\right)(\tau_m+1)
\exp \left(-c_2 \frac{\nmck
    \zeta_k^2}{\max_{i}q(y_i|\e_{k-1})^4}\right)\\
+\sup\limits_{\eta_{k-1} \in \mathcal{L}_m }\left(\sum_{i=1}^n
  \frac{1}{q(y_i|\e_{k-1})}\right) (\tau_m+1)\zeta_k.
\end{multline*}
We have to prove that the Monte Carlo sum involved in $
S_{\nmck}(s,y_i|\xi_{s,i},\e_{k-1})$ does not equal zero everywhere, so that $\sup_{\xi,s} S_{\nmck}(s,y_i|\xi_{s,i},\e_{k-1}) $ is finite.
For this purpose, we can choose a particular probability density
function $f$. Indeed, if we set $f$ to be the prior density function
on the simulated deformation fields $ \xi$, we have for all $\e \in \mathcal{V}_\m$:
\begin{eqnarray*}
\frac{1}{\nmc} \sum\limits_{l=1}^\nmc
 \left[
   \frac{\densMC(\xi^{(l)} _{t,i})}{q(y_i,\xi^{(l)} _{t,i},t|\e)}
 \right] &=& \frac{1}{\nmc} \sum\limits_{l=1}^\nmc
 \left[
   \frac{1}{q(y_i|\xi^{(l)} _{t,i},t,\e)q(t|\e)}
 \right]\\
&\geq & \frac{1}{\nmc} \sum\limits_{l=1}^\nmc
 \left[
   \frac{1}{
\frac{1}{(2\pi \si^2_t)^{|\Lambda|} }\exp\left(-\frac{1}{2\si^2_t} \|y_i -
K_p^{\xi^{(l)}} \alpha_t\|^2\right)
}\right]
\geq (2\pi \si^2)^{|\Lambda|},
 \end{eqnarray*}
where $\si$ is the lower bound of the variances $(\si_t)$.

We choose the sequence $(\zeta_k)_k$ depending upon $(\nmck)_k$ such that
$\lim\limits_{k\to\infty} \zeta_k=0$ and $\lim\limits_{k\to\infty} \nmck\zeta_k^2=+\infty$.
We can take for example $\zeta_k=J_k^{-1/3}$ for all
$k \geq 1$.
\vspace{0.3cm}

 We will now prove the convergence
 of the sequence of excitation terms.

For any $M>0$ we define  $M_n=\sum\limits_{k=1}^n \Delta_k
e_k\mathds{1}_{W(s_{k-1})\leq M}$ and let $\mathcal{F}=
 (\mathcal{F}_k)_{k\geq 1}$ be the filtration,
 where $\mathcal{F}_k$ is the $\si-$algebra generated by the random
 variables $(S_0,\bdbeta_1, \dots,
 \bdbeta_k,\bdtau_1,\dots,\bdtau_k)$.
We have
$M_n = \sum\limits_{k=1}^n \Delta_k \left(
\tS (\bdbeta_k,\bdtau_k)     - \mathbb{E} \left[
\tS (\bdbeta_k,\bdtau_k) |
\mathcal{F}_{k-1}
\right]
\right)\mathds{1}_{W(s_{k-1})\leq M}
$
so this shows us that $(M_n)$ is
a $\mathcal{F}$-martingale.
In addition to this we have:
\begin{eqnarray*}
\sum\limits_{k=1}^\infty \mathbb{E}\left[\|M_{k}-M_{k-1}\|^2 \left| \right.
\mathcal{F}_{k-1} \right] &= &
\sum\limits_{k=1}^\infty
\mathbb{E}\left[\Delta_{k}^2\|e_{k}\|^2\mathds{1}_{W(s_{k-1})\leq M}
  \ |\ \mathcal{F}_{k-1} \right]
\leq \sum\limits_{k=1}^\infty \Delta_{k}^2
\mathbb{E}\left[\|e_{k}\|^2
  \ |\ \mathcal{F}_{k-1} \right]\\
&\leq& \sum\limits_{k=1}^\infty \Delta_{k}^2
\mathbb{E}\left[\| \tS(\bdbeta_k,\bdtau_k)   -
\mathbb{E}\left[\tS (\bdbeta_k,\bdtau_k)  |\mathcal{F}_{k-1}\right]\|^2
  \ |\ \mathcal{F}_{k-1} \right]\\
&\leq & \sum\limits_{k=1}^\infty \Delta_{k}^2
\mathbb{E}\left[\| \tS(\bdbeta_k,\bdtau_k)  \|^2
  \ |\ \mathcal{F}_{k-1} \right] \, .
\end{eqnarray*}
We now evaluate this last integral term:
\begin{eqnarray*}
\mathbb{E}\left[\| \tS(\bdbeta_k,\bdtau_k)  \|^2
  \ |\ \mathcal{F}_{k-1} \right]&=& \sum_{\la} \int_{\R^N} \int
\| \tS(\bdbeta,\bdtau)   \|^2
\Pi^{\nmck}_{\e_{k-1},\bdtau}(\bdbeta_{0},\bdbeta)  \prod_{i=1}^n
p_{\nmck,\e_{k-1}} (\tau_i,\xi_{\tau_i,i},y_i)
Q(\xi_{\tau_i,i})d\xi_{\tau_i,i} d\bdbeta\\
&\leq& \left[\sum\limits_{\la} \int_{\R^N}  \| \tS(\bdbeta,\bdtau)  \|^2
  \Pi^{\nmck}_{\e_{k-1},\bdtau}(\bdbeta_{0},\bdbeta) d\bdbeta \right]\left[ \int
\Pi^{\nmck}_{\e_{k-1},\bdtau}(\xi_{0},\xi)d\xi \right] \,.
\end{eqnarray*}
The last term equals one and again we only need
to focus on the sufficient statistic which is not bounded
itself. Indeed $S_{3,t}(\bdbeta,\bdtau)$ for all $ 1\leq t \leq \tau_m $
so using
the fact that the function $V$ dominates this sufficient statistic, we obtain:
\begin{eqnarray*}
\mathbb{E}\left[\| \tS_{3,t}(\bdbeta_k,\bdtau_k)  \|^2
  \ |\ \mathcal{F}_{k-1} \right]
&\leq& \sum\limits_{\bdtau} \int_ {\R^N} \| \tS_{3,t}(\bdbeta,\bdtau)  \|^2
\Pi^{\nmck}_{\e_{k-1},\bdtau}(\bdbeta_{0},\bdbeta) d\bdbeta  \\
&\leq& C\sum\limits_{\bdtau} \int_{\R^N}   V(\bdbeta)  ^2
\Pi^{\nmck}_{\e_{k-1},\bdtau}(\bdbeta_{0},\bdbeta) d\bdbeta
\leq C\sum\limits_{\bdtau}   \Pi^{\nmck}_{\e_{k-1},\bdtau}  V(\bdbeta_0)^2\,.
\end{eqnarray*}
Applying Lemma \ref{lem:2} for $p=2$, we get:
\begin{eqnarray*}
\mathbb{E}\left[\| \tS(\bdbeta_k,\bdtau_k)  \|^2
  \ |\ \mathcal{F}_{k-1} \right]
&\leq& C\sum\limits_{\la} ( \rho  V(\bdbeta_0)^2 +C)
\leq C\tau_m^n( \rho  V(\bdbeta_0)^2 +C) \, .
\end{eqnarray*}
Finally it remains:
$\sum\limits_{k=1}^\infty \mathbb{E}\left[\|M_{k}-M_{k-1}\|^2 \left| \right.
\mathcal{F}_{k-1} \right]
\leq  C \sum\limits_{k=1}^\infty \Delta_{k}^2 \,,$
which ensures that  the previous series
converges. This  involves
 that $(M_k)_{k\in\mathbb{N}}$ is a martingale bounded in $L^2$ so that
 $\lim\limits_{k\to\infty} M_k$ exists (see
\cite{hallheyde}). This proves the first part of (\textbf{STAB2}).

 To conclude this proof we apply Theorem \ref{SA} and get that
 $\lim\limits_{k\to\infty} d(s_k,\mathcal{L}' ) =0  $.

\section{Conclusion and discussion}\label{conclusion}

We consider the setting of Bayesian non-rigid
deformable models building in the context of \cite{AAT} and the
associated MAP estimator.
We approximate this estimator of this generative model parameters thanks to a
stochastic algorithm  which derives from an EM algorithm. We also
prove its theoretical convergence toward a critical point of the observed
likelihood.
This is, to our best knowledge, the first convergent estimation
algorithm of the 
templates and geometrical variabilities in the framework of mixture
model for deformable templates.
The algorithm is based on a stochastic approximation  of
the EM algorithm using  a MCMC approximation of the posterior distribution and
truncation on random boundaries. We present experiments  on the
US-postal database  as well as on some 2D medical data. This
shows that the stochastic approach can be easily implemented with the
algorithm detailed here  and is
robust to noisy situations, giving better result than the previous
deterministic schemes.


Many interesting questions remain open.

The first goal of these model and algorithm is the estimation of some atlases
in a given population and the acceptable deformations of these
atlases that can explain the variability in the population. However,
this model, as soon as the parameters are estimated, can be used to
create a classifier. Given a new image, one can compute the most likely
component that this image belongs to. This computation requires to
evaluate the integral of the complete likelihood with respect to  the posterior
distribution as well as in the estimation process. A first proposition
to overcome this difficulty has been
given in \cite{AAT} while approximating the posterior distribution by
a Dirac on its mode. This gave very interesting results which are
presented in that paper. However, in the case of noisy images, the
same problem occurs and leads to bad classification ratios. Another
way has been proposed in \cite{aktmfca08} using the same methods as in this
paper, that is to say, using Monte Carlo Markov Chain methods. 
The results are impressive and the
improvement noticeable.

We have presented here a set of experiments on 2D images. The
generative model as well as the algorithm and the proof of its
convergence do not depend on the dimension of the images. The
implementation for 3D images is only a numerical issue. We are
currently working on the 3D codes to test this algorithm on real
medical databases.

An  interesting extension would be to consider diffeomorphic
mapping and not only displacement fields for the hidden
deformation. This appears to be particularly interesting in the context
of Computational Anatomy where a one to one correspondence between
the template and the observation is usually needed and cannot be
guaranteed with linear spline interpolation schemes. This extension
could be done in principle using tangent models based on geodesic
shooting in the spirit of \cite{vmty04}. Many numerical as well as
theoretical work need to be done in this area.

\bibliographystyle{abbrv}
\bibliography{bibcomp2}

\end{document}